\documentclass[useAMS,usenatbib]{mn2e}

\usepackage{epsfig}
\usepackage{deluxetable}

\def\h2{H$_2$}
\def\f0{$F_0$}

\newcommand\ion[2]{#1$\;${\small\rmfamily\@Roman{#2}}\relax}% 

\title[Average power density spectrum of {\em Swift} long GRBs]{Average power density spectrum of {\em Swift}
long gamma--ray bursts in the observer and in the source rest frames}
\author[C.~Guidorzi et~al.]{C.~Guidorzi,$^{1}$\thanks{E-mail:guidorzi@fe.infn.it}
R.~Margutti,$^{2}$ L.~Amati,$^{3}$  S.~Campana,$^{4}$ M.~Orlandini,$^{3}$ 
\newauthor P.~Romano,$^{5}$ M.~Stamatikos,$^{6,7}$ G.~Tagliaferri$^{4}$\\
\mbox{}\\
$^{1}$Department of Physics, University of Ferrara, via Saragat 1,
  I-44122, Ferrara, Italy\\
$^{2}$Harvard-Smithsonian Center for Astrophysics, 60 Garden St., Cambridge,
MA 02138, USA\\
$^{3}$INAF, Istituto di Astrofisica Spaziale e Fisica Cosmica, Bologna, Via Gobetti 101,
I-40129 Bologna, Italy\\
$^{4}$INAF, Osservatorio Astronomico di Brera, Via Bianchi 46, I-23807 Merate (LC), Italy\\
$^{5}$INAF, Istituto di Astrofisica Spaziale e Fisica Cosmica, Palermo, Via U. La Malfa 153,
I-90146, Palermo, Italy\\
$^{6}$NASA Goddard Space Flight Center, Greenbelt, MD 20771, USA\\
$^{7}$Department of Physics, Ohio State University, 191 West Woodruff Avenue, Columbus,
OH 43210, USA
}

\begin{document}

\date{\today}

%\pagerange{\pageref{firstpage}--\pageref{lastpage}} \pubyear{2002}

\maketitle

\label{firstpage}

\begin{abstract}
We calculate the average power density spectra (PDS) of 244 long gamma-ray
bursts detected with the {\em Swift} Burst Alert Telescope in the 15--150~keV
band from January 2005 to August 2011.
For the first time we derived the average PDS in the source
rest frame of 97 GRBs with known redshift.
For 49 of them an average PDS was also obtained in a common source-frame
energy band to account for the dependence of time profiles on energy.
Previous results obtained on BATSE GRBs with unknown redshift showed that
the average spectrum in the 25--2000~keV band could be modelled with a
power-law with a $5/3$ index over nearly two decades of frequency with a
break at $\sim$1~Hz.
Depending on the normalisation and on the subset of GRBs considered, our
results show analogous to steeper slopes (between $1.7$ and $2.0$) of the
power-law. However, no clear evidence for the break at $\sim$1~Hz was found,
although the softer energy band of BAT compared with BATSE might account for that.
We instead find a break at lower frequency corresponding to a typical source
rest frame characteristic time of a few~seconds.
We furthermore find no significant differences between observer and source
rest frames. Notably, no distinctive PDS features are found for GRBs with
different intrinsic properties of the prompt emission either.
Finally, the average PDS of GRBs at higher redshifts shows possibly
shallower power-law indices than that of low-$z$ GRBs. It is not clear whether
this is due to an evolution with $z$ of the average PDS.
\end{abstract}

\begin{keywords}
gamma-rays: bursts --- radiation mechanisms: nonthermal
\end{keywords}

%%%%%%%%%%%%%%%%%%%%%%%%%%%%%%%%%%%%%%%%%%%%%%
\section{Introduction}
\label{sec:intro}
%%%%%%%%%%%%%%%%%%%%%%%%%%%%%%%%%%%%%%%%%%%%%%
Within the prompt emission of long gamma-ray bursts (GRBs),
different degrees of variability are observed over time scales
spanning from milliseconds \citep{Bhat92,Walker00} up to several seconds.
For some GRBs, variability seems to be mostly concentrated on
either a unique or more distinct time scales: a fast component
characterised by sub-second variability, superposed to a
slow one which comprises the broad pulses and the overall
temporal structure \citep{Scargle98,Vetere06,Margutti09,Gao11}.
Owing to the variety of GRB time profiles, a simple
characterisation of GRB temporal variability is still missing.
Despite some claims on occasional events, no unambiguous
evidence for coherent pulsations has been found
(e.g., Cenko et al. 2010; De~Luca et al. 2010).\nocite{Cenko10,DeLuca10}
In principle, a better characterisation of variability can
help to constrain the radiation mechanism and dissipation
processes responsible for the burst itself, which is yet one
of the least understood aspects of the overall GRB phenomenon
(e.g., see the reviews by Ghisellini 2011; Zhang 2011).
\nocite{Ghisellini11,Zhang11}
Theoretical interpretations of the fast and slow components
have already been put forward: e.g., in the context of a
relativistic jet making its way out through the stellar envelope,
the slow component arises from the propagation through
the star, while the fast component keeps track of the
inner engine activity \citep{Morsony10}.
This could be the formation of an hyperaccreting black hole
(e.g., MacFayden \& Woosley 1999; Kawanaka \& Kohri 2011).
\nocite{MacFayden99,Kawanaka11}
Alternatively, in the internal-collision-induced magnetic
reconnection and turbulence (ICMART) model the inner engine would
be responsible for the slow variability, while the fast term is
due to relativistic magnetic turbulence taking place in the
region where the burst itself is produced \citep{ICMART}.

The study of the Fourier power density spectrum (PDS)
provides important clues on the timing properties of astrophysical
sources. Fast-Fourier transform (FFT) techniques allow to
discover features like periodic or quasi-periodic pulsations,
and to study the continuum or red noise connected with some
aperiodic correlated variability in the time series
(e.g., van der Klis 1989; Israel \& Stella 1996).\nocite{Klis89,Israel96}
Beloborodov, Stern, \& Svensson (1998; 2000; hereafter, BSS98 and BSS00)
\nocite{Beloborodov98,Beloborodov00}
studied the average PDS of hundreds of GRBs detected with BATSE
\citep{Paciesas99} in the 25--2000~keV energy band, and found that
it can be described by a power-law $f^{-\alpha}$ with $\alpha\sim5/3$
over almost two decades in frequency, from $\sim10^{-2}$ to $\sim1$~Hz.
Above 1~Hz the PDS has a sharp break and steepens significantly.
They also observed the same behaviour in the PDS of some
individual, long, and bright GRBs.
Similar results were also obtained on a smaller sample of GRBs
detected with INTEGRAL \citep{Ryde03}.
However, this investigation has never been done in the source
rest frame due to the poor number of GRBs with known distance
observed with past experiments. Instead, it is important to
perform this analysis in the source rest frame, because the observed
time profiles are affected by different effects, such as time
dilation \citep{Norris94}, and narrowing of pulses at different
energies \citep{Fenimore95,Norris96}, which do depend on redshift.
As a consequence, average temporal properties might be affected.
As suggested by BSS98, the $5/3$ power-law is observed in
hydrodynamics in the Kolmogorov spectrum of velocity
fluctuations within a medium characterised by fully
developed turbulence. Within the GRB outflow, the relativistic
aberration of light makes the radiation very sensitive to
the velocity direction of the emitting blob, and this could
link the velocity spectrum with the observed PDS
\citep{Narayan09}.

In the context of the internal shock model \citep{Rees94}, 
\citet{Panaitescu99} and \citet{Spada00} used the observed $5/3$
slope to constrain the parameters' space of the wind ejection features
which determine the dynamics and optical thickness of the wind
of relativistic shells colliding with each other.
In addition, assuming that GRB central engines are powered by
neutrino-cooled accretion flows, the expected variability can be
characterised by a power-law PDS with an index ranging between $1.7$
and $2.0$ in the $0.1$--$100$~Hz \citep{Carballido11}.

As also pointed out by BSS98, the operation
of averaging out hundreds of PDSs of different GRB light curves
relies on the assumption that different GRBs are different
realisations of the same stochastic process.
It is therefore important to study the distribution of the power
at each frequency bin for a given sample of GRBs, and to see
whether the nature of the distribution depends on frequency.

In this work we carry out an analogous study on the average PDS
of different sets of GRBs detected in the 15--150~keV energy band
with the Burst Alert Telescope (BAT; Barthelmy et al. 2005)
\nocite{Barthelmy05} aboard the {\em Swift} satellite \citep{Gehrels04}.
In addition to being an independent data set
obtained with a very different detector, unlike for previous GRB
catalogues a significant fraction ($\sim$~30\%) of {\em Swift} GRBs
have a measured redshift. This allowed us to carry out the same
investigation in the source rest frame, and to test whether
GRBs with different intrinsic properties of the prompt emission
also exhibit systematically different PDSs.
The paper is organised as follows: Section~\ref{sec:data}
describes the data selection and analysis. Results are presented
in Section~\ref{sec:res}, followed by a discussion in
Section~\ref{sec:con}.

%%%%%%%%%%%%%%%%%%%%%%%%%%%%%%%%%%%%%%%%%%%%%%%%%%%%%%%%%%%
\section{Data analysis}
\label{sec:data}
%%%%%%%%%%%%%%%%%%%%%%%%%%%%%%%%%%%%%%%%%%%%%%%%%%%%%%%%%%%

%--------------------------------------------------
\subsection{Data selection}
\label{sec:data_sel}
%--------------------------------------------------
We initially started with a sample of 582 GRBs detected and
covered by BAT in burst mode from January 2005 to August 2011.
All the ground-discovered GRBs were excluded to ensure
homogeneity of the sample of mask-weighted, background-subtracted
light curves with high-time resolution.
We selected the long bursts by requiring $T_{90}>3$~s, where
$T_{90}$ were taken from the corresponding BAT refined GCN circulars,
or from the second BAT catalogue \citep{Sakamoto11} when not available.
Short GRBs with long extended emission with $T_{90}>3$~s
\citep{Sakamoto11} were excluded.
After calculating the PDS for GRBs with different peak rates,
we rejected those with a poor signal-to-noise ratio (S/N) by setting a
threshold on the peak rate of $0.1$~count~s$^{-1}$ per fully illuminated
detector for an equivalent on-axis source.
Peak rates were calculated on variable time scales, from a minimum
value of 64~ms.
At this stage we ended up with 313 GRBs.
For these GRBs we extracted the 64-ms mask-weighted light curve
in the 15--150~keV energy band.
To this aim, the BAT event files were retrieved and
processed with the HEASOFT package (v6.11) following the
BAT team threads.\footnote{
http://swift.gsfc.nasa.gov/docs/swift/analysis/threads/bat\_threads.html}
Mask-weighted light curves were extracted using the ground-refined
coordinates provided by the BAT team for each burst through the
tool {\tt batbinevt}. The BAT detector quality map of each GRB was
obtained by processing the next earlier enable/disable map of the
detectors. The resulting light curves are expressed as background-subtracted
count rates per fully illuminated detector for an equivalent on-axis
source. The count rates are normally distributed, and the corresponding
$\sigma$'s are also derived as the relative uncertainties.
Consistent results are obtained by using the {\tt batgrbproduct} script.

Finally, each noise-subtracted PDS (Section~\ref{sec:pds_calc})
had its frequency bins grouped so as to fulfil a 3$\sigma$ criterion.
We finally excluded all the GRBs whose PDS consisted of
less than 4 grouped bins. We ended up with 244 GRBs in the observer
frame, which hereafter will be referred to as the full sample
(Table~\ref{tbl-1}).
For these GRBs we also extracted the light curves in two
energy channels: 15--50~keV and 50--150~keV.
The aim is to study how the PDS results compare at different
energies.

%--------------------------------------------------
\subsection{Subsamples with known redshift}
\label{sec:data_z}
%--------------------------------------------------
For 150 GRBs of the initial sample with $T_{90}>3$~s the redshift
$z$ is known. We therefore extracted the corresponding 15--150~keV
light curves by using a common source rest frame bin time of 8~ms.
We recalculated the peak rate of the source rest frame light curves
and applied the same filtering as above, i.e. a threshold to the peak
rate, and a minimum number of grouped frequency bins in the
noise-subtracted PDS.
We ended up with a sample of 97 GRBs with known redshift.
Hereafter, we refer to this as the $z$-silver sample
(Table~\ref{tbl-2}).

We also wanted to account for the different rest-frame energy passbands,
so we selected a common source-frame energy band as a result of
a trade-off between the need for a sufficiently broad band so as to
collect enough photons, and collecting as many GRBs as possible.
We therefore chose the 66--366~keV energy band (rest frame), which is
fully covered by the 15--150~keV BAT passband\footnote{We did not
consider photons above 150~keV because of the drop in the BAT effective
area for such energies \citep{Barthelmy05}.} for 62 GRBs with $1.4<z<3.5$.
For these GRBs we re-extracted the corresponding light curves in the
common source rest frame energy band with a common rest-frame bin time
of 4~ms. Applying the same filtering criteria above we ended up with
a final sample of 49 GRBs.
Hereafter, this will be referred as the $z$-golden sample
(Table~\ref{tbl-3}).

%
% ++++++++++++++ Figure 1 ++++++++++++++
\begin{figure}
\centering
\includegraphics[width=8cm]{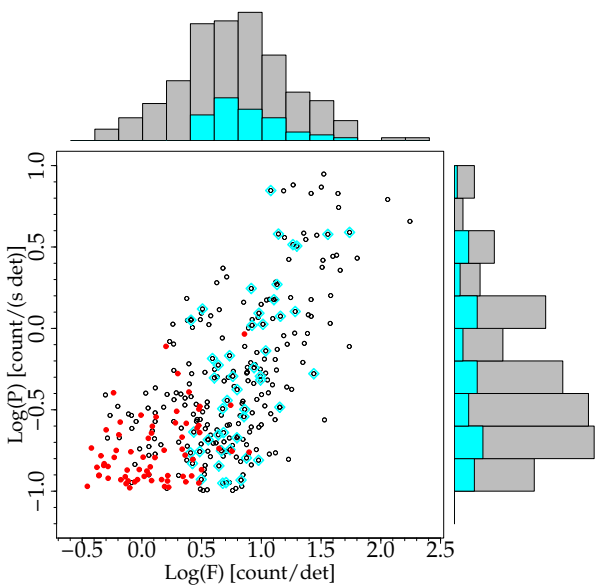}
\caption{The full set of 244 GRBs selected in the observer frame
(empty circles) and the $z$-golden subset with known $z$ and a common
source rest frame energy band (diamonds) shown in the peak rate-
count fluence plane (observer frame). Filled circles show the GRBs
discarded by our filtering procedure because of their poor PDS.
The corresponding histograms are also shown along both axes.}
\label{fig1}
\end{figure}
%++++++++++++++++++++++++++++++++++++++++++
%
Figure~\ref{fig1} shows the $z$-golden sample together with the full sample
of 244 GRBs in the $\log{P}$--$\log{F}$ plane evaluated in the observer
frame, where $P$ is the peak rate and $F$ is the count fluence per fully
illuminated detector for an on-axis source.
While the two samples have similar $P$ distributions,
that with known $z$ seems to be biased against low-fluence GRBs.
However, a Kolmogorov-Smirnov (K--S) test yields a $9.7$\% probability
for the two distributions of being drawn from the same population.
The $z$-golden sample thus is not inconsistent with being an unbiased
selection of the full set in the $\log{P}$--$\log{F}$ plane.
From Figure~\ref{fig1} it is also apparent that extrapolating to
peak rates below the threshold of $0.1$~count~s$^{-1}$~det$^{-1}$
would not increase the sample of GRBs with useful S/N, because
the fraction of rejected GRBs becomes dominant.

%--------------------------------------------------
\subsection{PDS calculation}
\label{sec:pds_calc}
%--------------------------------------------------
The choice of the time interval over which the PDS is most conveniently
calculated was driven by the need for covering the overall GRB profile
as well as optimising the S/N.
We verified that the overall shape of the PDS does not depend on the
particular choice of the time interval, whereas its S/N clearly does.
After a number of attempts, we came up with the following choice: first
we found the first ($t_1^{(n\sigma)}$) and the last ($t_2^{(n\sigma)}$) time
bins whose count rates exceeded the background level at $n$$\sigma$
($n\ge5$).
Let $\Delta_{n\sigma}=t_2^{(n\sigma)}-t_1^{(n\sigma)}$ be the duration of this interval.
The time interval chosen for the PDS calculation starts at
$t_1^{(n\sigma)}-\Delta_{n\sigma}$ and ends at $t_2^{(n\sigma)}+\Delta_{n\sigma}$.
If we had chosen a fixed time interval for all GRBs, thus with a common
frequency binning scheme, the S/N of the shortest GRBs would have been
worse, due to including additional noise with no signal.
In the observer frame the bin time was fixed to 64~ms, while for
the source rest frame GRBs two rest-frame bin times were used,
4 and 8~ms.
We found no noticeable difference between $n=5$ and $n=7$, apart from a
different S/N in the average PDS, which led us to finally choose $n=7$.
Tables~\ref{tbl-1}--\ref{tbl-3} report the time intervals
for all GRBs in each GRB sample.

In the full sample the duration of this time interval is found to be
thrice as long as the GRB duration expressed by its $T_{90}$. Given that
the time interval duration is by construction $3\,\Delta_{7\sigma}$, this
simply reflects that, on average, $\Delta_{7\sigma}$ does not differ
from $T_{90}$ remarkably.

The PDS was obtained through the mixed-radix FFT algorithm implemented
within the GNU Scientific Library
\citep{Galassi09},\footnote{http://www.gnu.org/s/gsl/} which does not
require the total number of bins to be a power of 2 \citep{Temperton83}.
Each PDS was calculated adopting the Leahy normalisation, in which the
constant power due to statistical noise has a value of 2 \citep{Leahy83}.
Usually the PDS is calculated from the light curves {\em not}
background-subtracted to ensure that counts are Poisson distributed,
and consequently the power distribution is known: e.g., in case of pure
statistical noise, the power is $\chi^2_2$-distributed.
In the case of BAT data, the background subtraction through the
mask weighting technique is not an issue as explained below. 

The average PDS of a given sample of GRBs was obtained assuming two different
normalisations: i) the BAT mask-weighted count rates and corresponding
errors are normally distributed and there is no evidence for any extra
variance (down to a few percent) in addition to the statistical white noise
\citep{Rizzuto07}. The uncertainties on the power of the individual
PDSs were calculated according to \citet{Guidorzi11}.
From Parseval's theorem the integral of each individual noise-subtracted
PDS yields the net variance, i.e. removed of the statistical noise.
In this case each PDS was normalised by its net variance. This
normalisation is preferable, because all GRBs have equal weights in the
average PDS.
ii) Each GRB light curve is normalised by its peak rate, which allows us
to make a direct comparison with BSS98 and BSS00.
Hereafter, the two cases are referred to as the
(net) variance and the peak normalisations, respectively.
A possible third normalisation based on the count fluence was soon
neglected, due to the results on the binned average PDSs, which were
statistically poorer and less constraining than for the other
normalisations. Using different normalisations allows us to evaluate
the effects of this kind of choice.

The statistical noise was removed differently in the two cases:
in i) the noise was assumed to be perfectly Poissonian \citep{Rizzuto07},
and calculated consequently. In ii) it was obtained from fitting the
average PDS with a constant at sufficiently high frequencies.

We started from a uniform frequency binning scheme with a step of
$0.01$~Hz. At $f<0.01$~Hz we considered two bins,
$0.001$~Hz~$\le f<0.005$~Hz and $0.005$~Hz $\le f<0.01$~Hz.
For each frequency bin and for each individual GRB we calculated
the average power. Finally, for each frequency bin we averaged out the
power over all the GRBs of a given sample. The average power in each bin
is approximately normally distributed with $\sigma=\sigma_{\rm p}/\sqrt{n}$,
where $\sigma_{\rm p}$ is the standard deviation of the corresponding
power distribution and $n$ the size of the array, i.e. roughly
the size of a given GRB sample.
Its validity is ensured by the central limit theorem.
$\sigma$ was therefore taken as the uncertainty of the
corresponding average power.
Finally the frequency bins of the average noise-subtracted PDS
were grouped by requiring at least 3$\sigma$ significance.

%--------------------------------------------------
\subsection{PDS modelling}
\label{sec:pds_fit}
%--------------------------------------------------
We modelled the average PDSs with a smoothly
broken power-law as in equation~(\ref{eq:mod}),

\begin{equation}
{\rm PDS}(f) = 2^{1/n}\,F_0\ \Big[ \Big(\frac{f}{f_{\rm b}}\Big)^{n\,\alpha_1} +
\Big(\frac{f}{f_{\rm b}}\Big)^{n\,\alpha_2} \Big]^{-1/n}
\qquad ,
\label{eq:mod}
\end{equation}
where the following parameters were left free to vary: the break frequency
$f_{\rm b}$, the value of the PDS at the break frequency $F_0$, the
two power-law indices $\alpha_1$ and $\alpha_2$ ($\alpha_2>\alpha_1$).
Initially, the peakedness parameter $n$ was also left free to
vary. However in most cases the data were not sensitive to it,
so we fixed $n=10$ for all cases to ensure a more homogeneous
comparison between the best-fit values obtained over different sets.
This choice implies a rather sharp break around $f_{\rm b}$.

In Section~\ref{sec:peakedness} we discuss more in detail how much the
results, especially $\alpha_2$, depend on this degree of freedom.
The best-fit model is obtained by minimising the total $\chi^2$.

%%%%%%%%%%%%%%%%%%%%%%%%%%%%%%%%%%%%%%%%%%%%%%%%%%%%%%%%%%%
\section{Results}
\label{sec:res}
%%%%%%%%%%%%%%%%%%%%%%%%%%%%%%%%%%%%%%%%%%%%%%%%%%%%%%%%%%%
Table~\ref{tbl-4} reports the best-fit parameters of the model in
equation~(\ref{eq:mod}) for each GRB sample and both normalisations.
Parameters' uncertainties are given at 90\% confidence level for
one parameter of interest.

%
% ++++++++++++++ Figure 2 ++++++++++++++
\begin{figure}
\centering
\includegraphics[width=8cm]{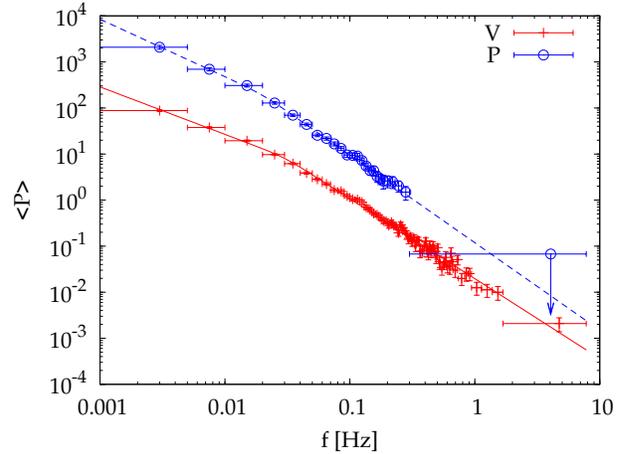}
\caption{Average PDS for the full sample of 244 GRBs in the
observer frame. Crosses (empty circles) correspond to the net variance
(peak rate) normalisation case. Upper limits are 2$\sigma$.
The peak rate data have conveniently been shifted for the sake of clarity.
The corresponding best fit broken power-laws are also shown.
Notably, the peak normalisation has a poorer S/N due to the bright GRBs
having a smaller weight than in the variance normalisation;
this becomes apparent at high frequencies,
where the signal becomes comparable with statistical noise.}
\label{fig2}
\end{figure}
%++++++++++++++++++++++++++++++++++++++++++
%
Figure~\ref{fig2} displays the average PDS of the full sample for
both normalisations as well as their corresponding best-fit models.
The variance-normalised PDS has a best-fit value around $1.03\pm0.05$ for
$\alpha_1$, followed by a break around $3\times10^{-2}$~Hz, above
which the slope becomes $\alpha_2=1.73_{-0.03}^{+0.04}$.
The peak-normalised PDS has a similar value for $f_{\rm b}$, and
steeper values for the the power-law indices: $\alpha_1=1.25_{-0.12}^{+0.11}$
and $\alpha_2=1.90_{-0.06}^{+0.07}$.
The difference $(\alpha^{\rm (peak)}_i-\alpha^{\rm (var)}_i)$ in the power-law
indices between the two normalisations is found to be in the range
$0.1$--$0.2$ for all the GRB subsamples considered, although it
is always compatible with zero at $3$$\sigma$.

%
% ++++++++++++++ Figure 3 ++++++++++++++
\begin{figure}
\centering
\includegraphics[width=8cm]{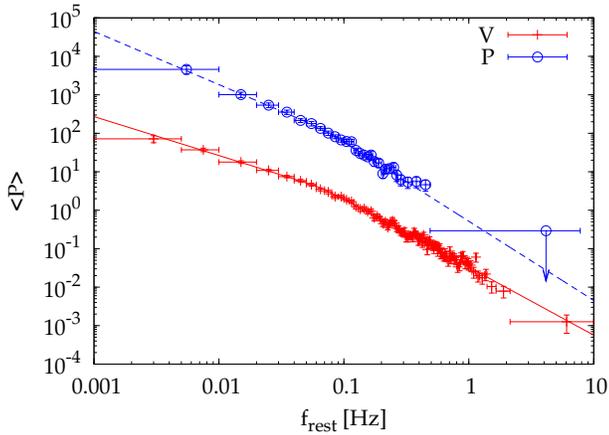}
\caption{Average source rest frame PDS for the $z$-silver sample of 97 GRBs 
in the observed 15--150~keV energy band. Symbols are the same as in
Fig.~\ref{fig2}.}
\label{fig3}
\end{figure}
%++++++++++++++++++++++++++++++++++++++++++
%
%
% ++++++++++++++ Figure 4 ++++++++++++++
\begin{figure}
\centering
\includegraphics[width=8cm]{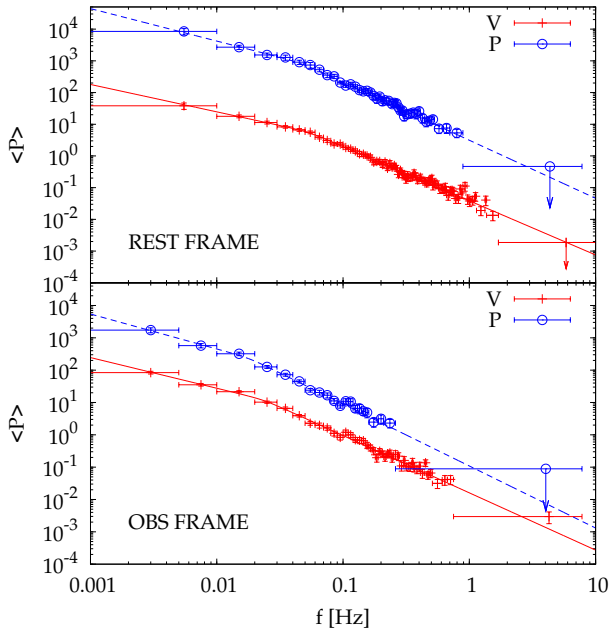}
\caption{{\em Top Panel}: average source rest frame PDS for the $z$-golden
sample in the common 66--366~keV source rest frame energy band.
{\em Bottom Panel}: average PDS for the same sample in the 15--150~keV energy
band (observer frame). Symbols are the same as in Fig.~\ref{fig2}.}
\label{fig4}
\end{figure}
%++++++++++++++++++++++++++++++++++++++++++
%
The value of $\alpha_2$ is very similar to the power-law index $1.67\pm0.02$
in the range $0.02<f<1$~Hz in the 50--300~keV band found by BSS98, and
between $1.50$ and $1.72$ in the range $0.025<f<1$~Hz in the 20--2000~keV
band found by BSS00.
Very similar results are obtained for the $z$-silver and golden samples
(Figs.~\ref{fig3} and \ref{fig4}),
apart from the best-fit values of $f_{\rm b}$ which are higher in the source
rest frame. This is no wonder, and links to the cosmological time dilation.
This is quantified by fitting the average PDS of the $z$-golden sample
in the observer frame: moving from observer to source rest frame,
$f_{\rm b}$ changes from $2.4\times10^{-2}$ to $5.3\times10^{-2}$~Hz.
The ratio of $2.2\pm0.6$ between the source- and the observer-frame
values of $f_{\rm b}$ is close to the average factor
of $(1+\hat{z})\sim3.2$, where $\hat{z}=2.2$ is both median
and mean redshift of the $z$-golden sample. 
Except for $f_{\rm b}$, the comparison of the results
obtained for the $z$-golden sample between observer and source rest frame
shows no significant differences in the power-law indices.
The same conclusion holds when we compare the $z$-silver with the full
sample. This result is not obvious: although power-law indices are
clearly invariant observables, the impact of averaging out different
source-frame energy bands on the observed PDS is not obvious.

%
% ++++++++++++++ Figure 10 ++++++++++++++
\begin{figure}
\centering
\includegraphics[width=8cm]{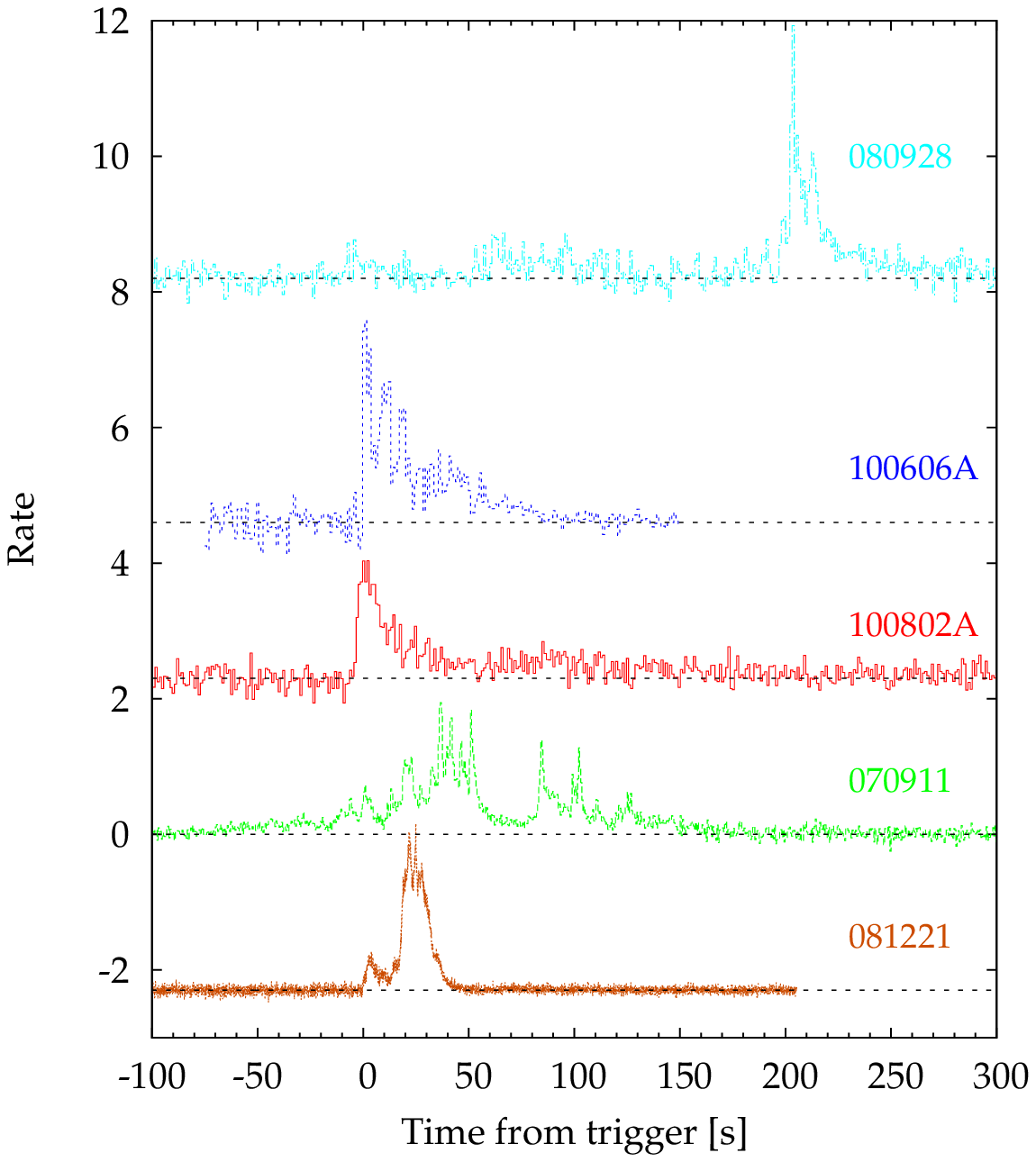}
\includegraphics[width=8cm]{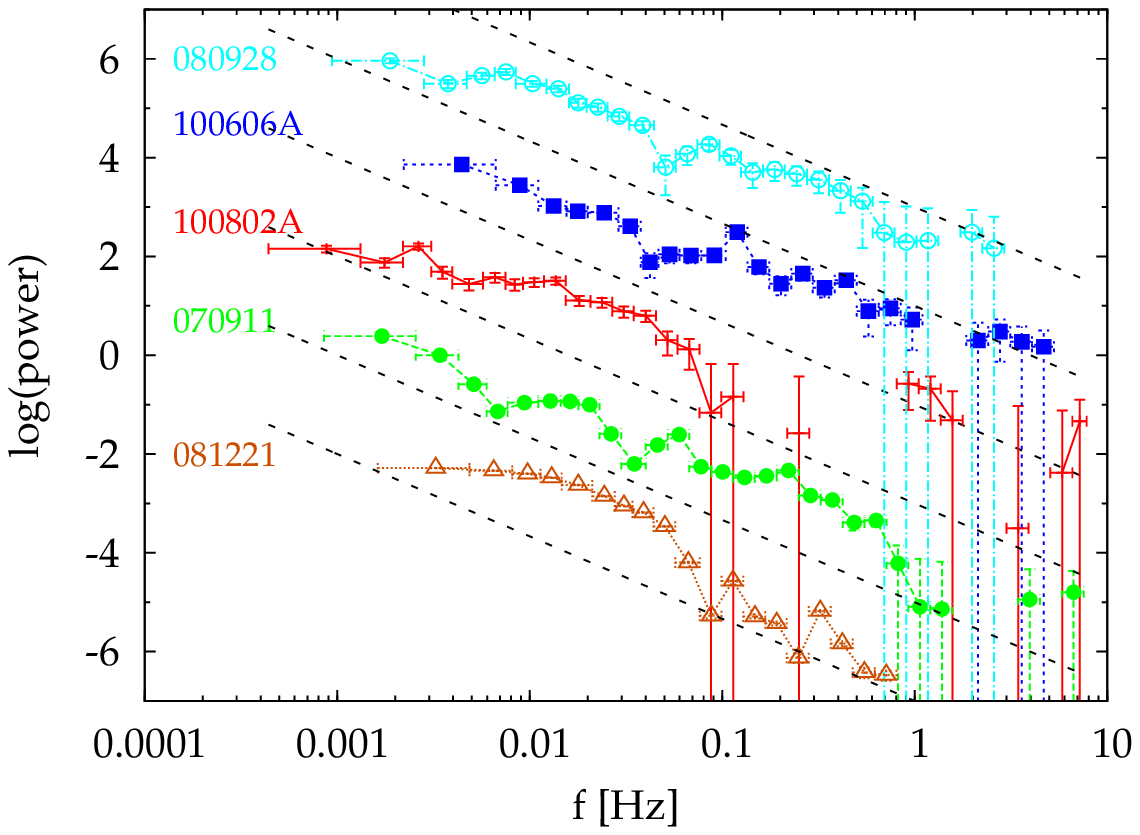}
\caption{{\em Top panel}: examples of light curves of individual GRBs randomly
selected from the full sample. Data are shifted along the $y$-axis for clarity.
{\em Bottom panel}: the corresponding PDSs. Dashed lines are power-laws with
index $5/3$.}
\label{fig10}
\end{figure}
%++++++++++++++++++++++++++++++++++++++++++
%
Figure~\ref{fig10} shows examples of light curves and their PDSs of
individual GRBs randomly picked out from the full sample.
While the global trend suggested by visual inspection favours the description
of the average PDS with a power-law with index compatible with $\sim1.7$
above a few~$10^{-2}$~Hz, individual PDSs still exhibit a variety of
different average declines.

%--------------------------------------------------
\subsection{Average PDSs at different energies}
\label{sec:res_en}
\begin{figure}
\centering
\includegraphics[width=8cm]{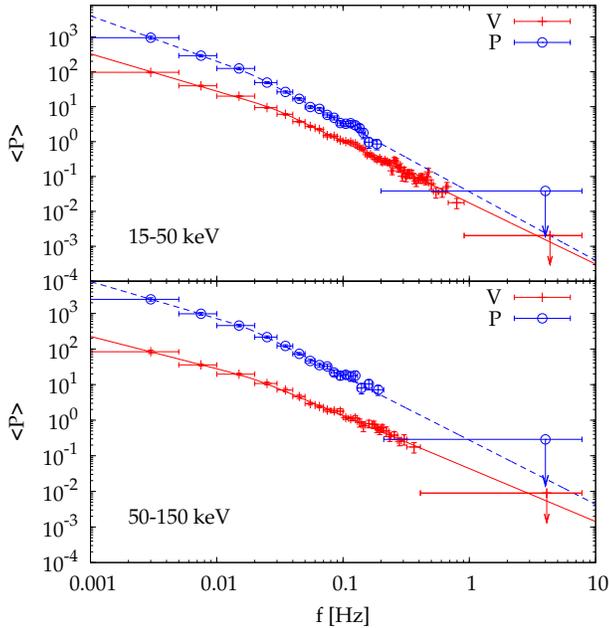}
\caption{{\em Top Panel}: average PDS for the full sample in the
15--50~keV energy band (observer frame).
{\em Bottom Panel}: average PDS for the same sample in the 50--150~keV
energy band. Symbols are the same as in Fig.~\ref{fig2}.}
\label{fig5}
\end{figure}

%--------------------------------------------------
Remarkably different behaviours in the high-frequency power-law indices are
observed for different observed energy bands: in the variance normalisation,
$\alpha_2$ varies from $1.75_{-0.04}^{+0.05}$ to $1.49_{-0.07}^{+0.08}$ passing
from 15--50 to 50--150~keV.
The low-frequency index $\alpha_1$ is also shallower at higher energies,
$0.91_{-0.14}^{+0.11}$ to be compared with $1.07\pm0.05$ observed in the
softer energy channel (Fig.~\ref{fig5}).
%
% ++++++++++++++ Figure 5 ++++++++++++++
%++++++++++++++++++++++++++++++++++++++++++
%
The break frequency shows no significant dependence on energy.
Analogous variations are observed in the peak normalisations, although
the indices are systematically steeper, as noted above.
The same trend was noted in the individual BATSE energy channels:
the power-law index decreased from $1.72$ in the 25--55~keV to $1.50$
above 320~keV (BSS00).

%
% ++++++++++++++ Figure 11 ++++++++++++++
\begin{figure}
\centering
\includegraphics[width=8cm]{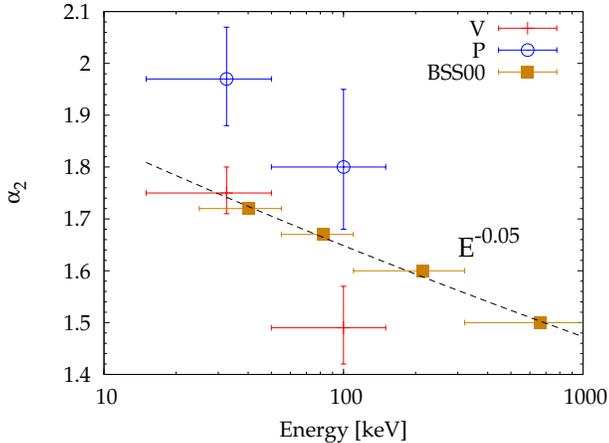}
\caption{High-frequency power-law index $\alpha_2$ as a function of
energy $E$ obtained by fitting the average PDS of the full sample
within the variance (crosses) and the peak (circles) normalisations.
BSS00 results derived from BATSE GRBs are also shown (squares).
The dashed line shows the power-law $\alpha_2\propto E^{-0.05}$.}
\label{fig11}
\end{figure}
%++++++++++++++++++++++++++++++++++++++++++
%
Figure~\ref{fig11} directly compares our values with BSS00's
as a function of energy. The BATSE data show a dependence on energy
$E$ which can be modelled with the power-law $\alpha_2\propto E^{-0.05}$.
These results are compatible with ours within uncertainties (BSS00
do not provide uncertainties on their values).
Curiously, the variance normalisation, which is preferable to us
also for the reasons explained below in Section~\ref{sec:res_dist}, is
in better agreement with BSS00 results than the peak normalisation,
which BSS00 adopted for their analysis.

%--------------------------------------------------
\subsection{The effects of GRB durations}
\label{sec:res_dur}
%--------------------------------------------------
The cut-off frequency $f_{\rm b}$ is mainly connected with the average
duration of the GRBs and of the individual pulses they consist of,
as noted above when moving from the observer to the source rest frame.
We investigated the role of the duration by selecting two subsets of
the full sample with extreme durations, each collecting 90 GRBs.

In principle, because of the finiteness of the GRB duration,
and therefore of the temporal window which contains its time profile,
the observed PDS is the convolution of the true one with
$|\sin{(f\,\pi\,T)}/(f\,\pi)|^2$, where $f$ is frequency and $T$
is the length of the window (e.g., van der Klis 1989).\nocite{Klis89}
In practice, all the features in the true PDS narrower than $1/T$ are
smoothed out, and the minimum frequency that can be explored, $1/T$,
also defines the resolution of the PDS. Provided that $T\gg\tau$, where
$\tau$ is a generic characteristic time scale acting in a GRB, the
cut-off frequency associated to it, $\approx1/\tau$, is unaffected.

\begin{figure}
\centering
\includegraphics[width=8cm]{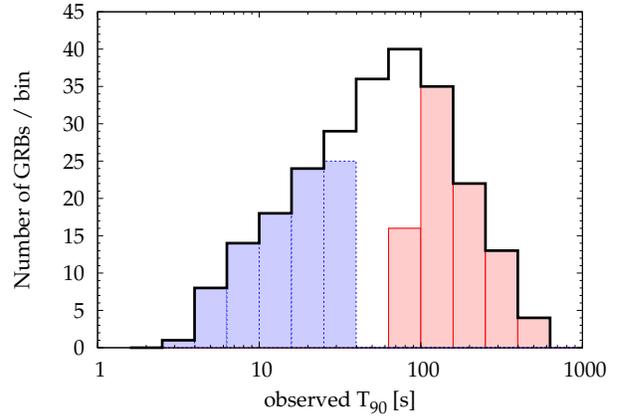}
\caption{$T_{90}$ distribution for the full sample (solid line). Shaded
histograms show two subsets including 90 GRBs each with the shortest and longest
durations, respectively.}
\label{fig6}
\end{figure}

\begin{figure}
\centering
\includegraphics[width=8cm]{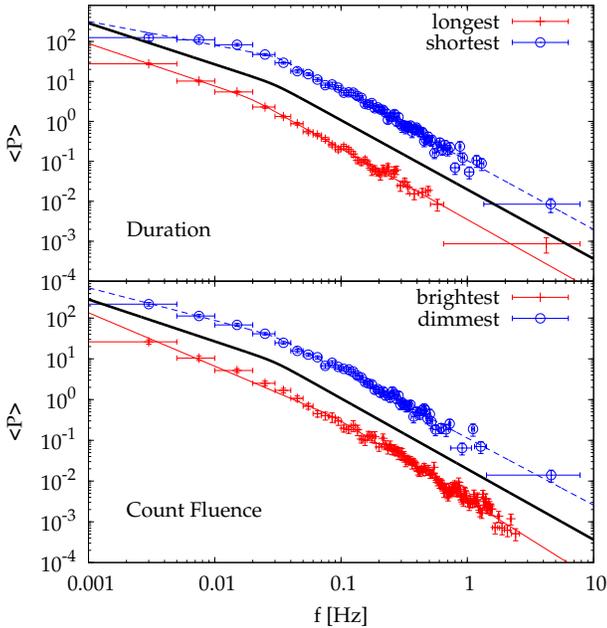}
\caption{{\em Top Panel}: average PDS for two subsets with extreme
durations: 90 shortest vs. 90 longest duration GRBs. 
{\em Bottom Panel}: average PDS for two subsets  with extreme
count fluences: 30 highest-fluence vs. 97 lowest-fluence GRBs.
The thick solid lines is the best fit for the full sample. Different data
sets are shifted for clarity reasons.}
\label{fig7}
\end{figure}

Figure~\ref{fig6} shows the duration distributions for the full sample
as well as for the two subsets chosen by us: GRBs with $T_{90}<40$~s,
$T_{90}>80$~s on the other side.
The top panel of Figure~\ref{fig7} displays the average PDS of both
subsamples.
%
% ++++++++++++++ Figure 6 ++++++++++++++
%++++++++++++++++++++++++++++++++++++++++++
%
%
% ++++++++++++++ Figure 7 ++++++++++++++
%++++++++++++++++++++++++++++++++++++++++++
%
The logarithmic average durations of each subsample are $15$ and $153$~s,
respectively. Indeed, $f_{\rm b}$ decreases from $3.2\times10^{-2}$ of the
shortest GRBs to $2.0\times10^{-2}$~Hz of the longest ones (Table~\ref{tbl-4}).
The change of $f_{\rm b}$ is not comparable to the corresponding change in the average
duration. However, the low-frequency power-law index $\alpha_1$ significantly
steepens from $\sim0.6$ to $\sim1.1$ in the variance normalisation.
This suggests that different values of $f_{\rm b}$ may affect the estimate
of $\alpha_1$, because the asymptotic behaviour is not reached when $f_{\rm b}$
is close to the lowest explorable frequency, and is more affected by the finite
width at low frequencies.
Moreover, the cut-off frequency is more sensitive to the average characteristic
times (both rise and decay times) of individual shots, rather than to the overall
duration: e.g., the PDS of a simple exponential shot with a characteristic
time $\tau$ has $f_{\rm b}\sim 1/2\pi\tau$ (e.g., Lazzati 2002).\nocite{Lazzati02}
This is still true in the presence of shot noise \citep{Frontera79,Belli92},
provided that the occurrence times of the shots are independently distributed.
Although this is a rough approximation in this case, the best-fit values
of $f_{\rm b}$ imply characteristic times of individual shots about 5--8~s
(2--4~s) long in the observer (source rest) frame.

%--------------------------------------------------
\subsection{The effects of GRB peak rates and fluences}
\label{sec:res_flu}
%--------------------------------------------------
Similarly, we investigated the effects of both the peak count rate and
fluence on the average PDS by selecting proper subsets of the full sample.
Unlike for the duration, the S/N does depend on both peak rate and fluence.
We ensured comparable statistical quality of the two subsets by collecting more
faint bursts (both in terms of peak rates and count fluence).
As for the peak rate $p$, we ended up with 124 and 65 GRBs with
$p<0.4$~count~s$^{-1}$~det$^{-1}$, and $p>1.0$~count~s$^{-1}$~det$^{-1}$,
respectively. The fluence $F$-selected subsets include 97 and 30 bursts with
$F<4.4$~count~det$^{-1}$, and $F>19.8$~count~det$^{-1}$, respectively.
Both peak rate and count fluence distributions are shown in the projected
histograms of Fig.~\ref{fig1}.

Concerning the peak rate, the best-fit power-law indices are the same for both
subsamples. Instead, $f_{\rm b}$ increases by a factor of 3 (5) in the variance
(peak) normalisation when we move from the faint to the bright subset.
This is due to the brighter pulses being narrower \citep{Norris96}: on average
the GRBs of our faint subset have 3 to 5 times longer pulses than those of the
bright subset.
This seems to be at variance with the results by BSS00, who found a decreasing
power-law index with increasing peak count rate: from $1.82$ for the faintest
GRBs down to $1.63$ for the brightest end.

Concerning $\alpha_1$, only for the count fluence sample this index becomes
shallower when moving from the high- to low-fluence subset.
The same behaviour is observed for the duration-driven subsamples
(Section~\ref{sec:res_dur}) when we move from the longest to the shortest GRBs.
This common property is explained by the shortest GRBs having lower fluence
on average, as confirmed by the correlation between fluence and $T_{90}$
for both the full and the $z$-silver samples with significance values of
the order of $10^{-11}$ and $0.1$--$0.2$\%, respectively, according to the
non-parametric tests of Spearman and Kendall.

We note that the estimate of the power-law index within a given range can
be affected by the choice of the model: from fig.~8 of BSS00, the range over which
the power-law is fitted extends over a single decade. Within such a limited range
the fitted slope is sensitive to the frequency interval chosen for
modelling. Within our data, when we opt for a smoother break in equation~(\ref{eq:mod})
by fixing $n=1$, the best-fit values for the post-break frequency slope $\alpha_2$
are systematically steeper, because the asymptotic value is not reached within
the range covered by the data (Section~\ref{sec:peakedness}).

In equation~(\ref{eq:mod}) this may introduce an artificial correlation between
$f_{\rm b}$ and $\alpha_2$ (as well as $\alpha_1$): the higher $f_{\rm b}$, the
steeper $\alpha_2$, when the frequency range covered by the data is not
sufficiently broad. Indeed, we observe this in the fluence-driven subsets,
as reported in Table~\ref{tbl-4} and shown in the bottom panel of Fig.~\ref{fig7}.
Only in the variance normalisation case, $\alpha_2$ varies from $1.65$ (faint subset)
to $1.94$ (bright subset), whereas $f_{\rm b}$ increases by a factor of $3.1$.
No such change is observed in the peak normalisation case, where both $f_{\rm b}$
and $\alpha_2$ experience very little changes.

We checked whether the results obtained on the fluence-driven subsets are affected
by the corresponding average S/N. We split the sample in two subsets of 50 and 150
GRBs having the highest- and lowest-S/N PDSs, respectively. Within uncertainties
the two average PDSs showed no distinctive behaviour. Therefore, the variety of
S/N is not directly responsible for the different PDS properties of
fluence-driven subsets.

%--------------------------------------------------
\subsection{The effects of different $E_{\rm iso}$ and $L_{\rm p,iso}$}
\label{sec:res_eiso}
%--------------------------------------------------
Here we investigate the possible existence of correlations between the PDS
and intrinsic properties of the prompt emission.
Out of the $z$-silver sample, we selected the GRBs with measured intrinsic
peak energy of the time-averaged $E\,F(E)$ spectrum, the so-called $E_{\rm p,i}$.
The isotropic-equivalent gamma-ray released energy in the $1$--$10^4$~keV band,
$E_{\rm iso}$, which correlates with $E_{\rm p,i}$ \citep{Amati02},
is also known for the same events.
We adopted the standard cosmological model: $H_0=71$~km~s$^{-1}$~Mpc$^{-1}$,
$\Omega_\Lambda=0.73$, $\Omega_M=0.27$ \citep{Spergel03}.
The values for both $E_{\rm p,i}$ and $E_{\rm iso}$ are taken from
Amati et al. (2008; 2009; in prep.).\nocite{Amati08,Amati09}
We ended up with a subset of 64 GRBs. For the same bursts we estimated the
isotropic-equivalent peak luminosity,
$L_{\rm p,iso}$, by normalising $E_{\rm iso}$ through the ratio
between peak count rate and count fluence in the source-rest frame.
This implicitly assumes no spectral evolution throughout the prompt emission,
which for instance does not hold when the hardness ratio tracks the time profile.
This leads to underestimating the peak luminosity by a factor of a few in the worst case.
However, in a logarithmic space this cannot wash out possible genuine
correlations or build fake ones, but it may merely increase the observed dispersion.
The values of $E_{\rm iso}$, $E_{\rm p,i}$, and $L_{\rm p,iso}$ for this subsample
are reported in Table~\ref{tbl-5}.

We compared the average PDS of the least and that of the most energetic GRBs
as follows:
we collected two subsets of 25 GRBs each, with extreme values of $E_{\rm iso}$.
Each subset collects about 1/3 of the overall set. The least energetic GRBs
all have $E_{\rm iso}<9\times10^{52}$~ergs, while the most energetic ones all
have $E_{\rm iso}>2.1\times10^{53}$~ergs, with logarithmic average values of
$2.8\times10^{52}$ and $4.7\times10^{53}$~ergs, respectively.
The average PDS of the two groups are not found to significantly differ from
one another, as reported in Table~\ref{tbl-4}.
%The average PDS of the most
%energetic GRBs shows the same trend as that of the GRBs with the largest fluence
%(Section~\ref{sec:res_flu}), i.e. higher values of both $f_{\rm b}$ and
%$\alpha_2$ with respect to the least energetic sample. However in this case
%the differences are milder and not statistically significant, possibly because
%of the poorness of the sample with known intrinsic quantities.
%We conclude that there is no clear evidence for a significantly different
%average PDS between the least and the most energetic bursts.

Analogously, we divided the same sample according to different classes of
$L_{\rm p,iso}$ and in none of the cases we found evidence for a dependence of
the average PDS on $L_{\rm p,iso}$.

%--------------------------------------------------
\subsection{The effects of redshift}
\label{sec:res_z}
\begin{figure}
\centering
\includegraphics[width=8cm]{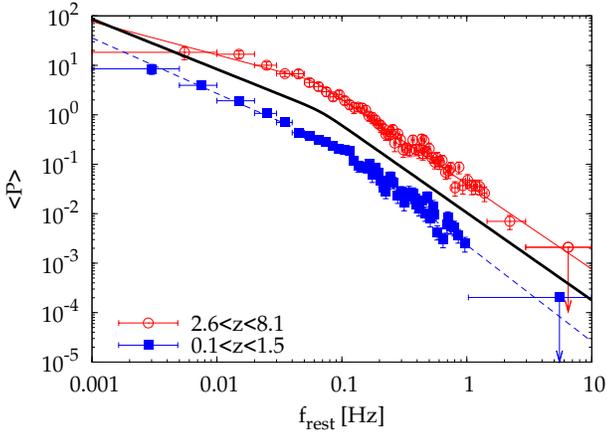}
\caption{Average source-rest frame PDS for two subsets with different
redshift bins taken from the $z$-silver sample: low-$z$ (filled squares),
and high-$z$ (empty circles), together with their corresponding best-fit
models. For comparison, the thick solid line shows the best-fit model
obtained over the entire $z$-silver sample.}
\label{fig12}
\end{figure}

\begin{figure}
\centering
\includegraphics[width=8cm]{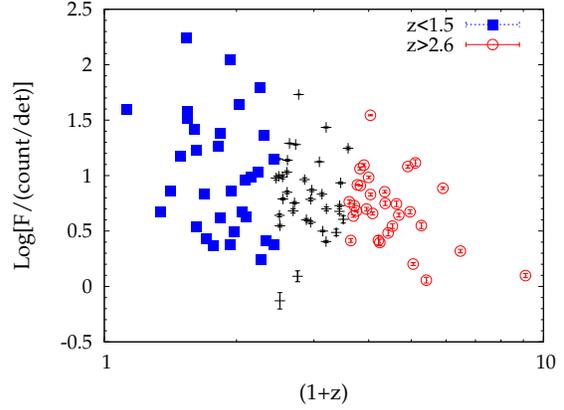}
\caption{Count fluence vs. redshift $z$ distribution for the $z$-silver sample.
The two subsets with different redshift bins are also shown with the same
symbols as in Fig.~\ref{fig12}.}
\label{fig13}
\end{figure}

%--------------------------------------------------
We considered two classes of 32 GRBs each with the lowest and highest
redshift values, respectively, among the $z$-silver sample.
The aim is to study possible evolutionary effects.
The choice of the number of GRBs is a trade-off between the need for
a big enough sample for statistical purposes, and the need of having two
well separated redshift bins. We came up with two classes: the low-$z$
GRBs with $0.1<z<1.5$, and the high-$z$ ones with $2.6<z<8.1$.
The mean (median) redshifts for both subsets are $0.9$ ($0.9$) and
$3.5$ ($3.2$), respectively.
The corresponding average PDSs are shown in Figure~\ref{fig12} together
with their best-fit models (Table~\ref{tbl-4}).
%
% ++++++++++++++ Figure 12 ++++++++++++++
%++++++++++++++++++++++++++++++++++++++++++
%
Similarly to what is observed for the low-fluence GRBs, for the variance
normalisation the average PDS of high-$z$ GRBs has shallower indices:
$\alpha_1$ and $\alpha_2$ are
respectively $0.67_{-0.22}^{+0.17}$ and $1.71_{-0.07}^{+0.08}$, to be compared
with $1.12\pm0.08$ and $1.95_{-0.11}^{+0.12}$ found for the low-$z$ GRBs.
To understand whether this is due to the farther GRBs having lower fluence
values on average, in Figure~\ref{fig13} we studied the correlation between
fluence and redshift for the full $z$-silver sample as well as for
the two subsets here considered.
%
% ++++++++++++++ Figure 13 ++++++++++++++
%++++++++++++++++++++++++++++++++++++++++++
%
There is a hint for an anticorrelation between observed fluence and redshift,
whose significance is about $0.2$--$0.3$\% according to non-parametric tests
(Spearman, Kendall).
The fluence distributions of the low- and of the high-$z$ subsets have a
K--S probability of $1.6$\% of being drawn from the same population.
As for the fluence, the $z$-silver sample is an unbiased subset of the full
sample, since the two fluence distributions are fully compatible
(46\% probability according to a K--S test). Thus, the low-fluence subset of
the full sample discussed in Section~\ref{sec:res_flu} is likely to include
more high-$z$ bursts than what the high-fluence subset does.
On average, the high-fluence bursts have lower redshifts and this explains
the common properties observed in the average PDS, compared with that of
low-fluence and high-redshift GRBs.
Given the correlation between fluence and $T_{90}$, we checked whether
for the $z$-silver sample the redshift also correlates with the observed
$T_{90}$, and we found it does not. Therefore, while the fluence correlates
with $T_{90}$ and anticorrelates with redshift, the latter does not correlate
with $T_{90}$.

Qualitatively, the shallower power-law indices for the low-fluence/high-$z$
GRBs can be explained by the result found on the average PDSs of different
energy channels (Section~\ref{sec:res_en} and Fig.~\ref{fig11}): harder
photons have a shallower PDS (see also Table~\ref{tbl-4}).
Given that the light curves of the $z$-silver
sample refer to the common {\em observed} 15--150~keV energy band, the
results obtained for the high-$z$ subset refer to a harder source-rest frame
energy band.
Whether this difference between the average PDS of low- and that of
high-$z$ GRBs can entirely be ascribed to the cosmological shift of the
energy band, or it is due to an evolutionary property of GRBs is not clear.
To clarify this issue, we should apply the same analysis to
the $z$-golden sample: however, practically this is not feasible due to
the low number of GRBs, the limited range both in $z$ (Section~\ref{sec:data_z})
and in fluence (Fig.~\ref{fig1}).

%--------------------------------------------------
\subsection{Smoothness of the break}
\label{sec:peakedness}
%--------------------------------------------------
In Section~\ref{sec:res_flu} we noted that choosing a smoother break in
equation~(\ref{eq:mod}), e.g. $n=1$ instead of $n=10$, yields
systematically shallower (steeper) values for $\alpha_1$ ($\alpha_2$).
The differences in $\alpha_2$ between $n=10$ and $n=1$ are however
milder and in all cases they are compatible with zero within uncertainties.
Analogously, $f_{\rm b}$ is systematically lower for a smoother break,
although not significantly. Overall, allowing the data to be fitted with
a smoother break implies that the asymptotic regime at low frequencies
is not covered by the data, and this leads to shallower values for
$\alpha_1$ than what data actually exhibit. The goodness of the fits in both
cases is similar and shows no systematic behaviour.

We conclude that the degree of freedom brought in by the smoothness of the
break of equation~(\ref{eq:mod}) does not affect significantly
the estimates of $\alpha_2$, thanks to the broader frequency range
at $f>f_{\rm b}$ covered by the data.

%--------------------------------------------------
\subsection{PDS distribution}
\label{sec:res_dist}
%--------------------------------------------------
PDSs of individual GRBs are very different from each other
(Fig.~\ref{fig10}). For instance,
let us consider a single fast-rise exponential decay (FRED) with a
characteristic time $\tau$ (either rise or decay time). At $f\gg1/\tau$
the PDS asymptotically declines as a power-law with an index of 2 or steeper.
Depending on the peakedness value, the PDS can also exhibit oscillatory
terms in the decay, as shown by \citet{Lazzati02}.
Oscillations modulating the power-law decline can also appear in the PDS
of those GRBs with two or more pulses separated by a quiescent time,
which makes them interfere in the Fourier transform (since the PDS can
also be seen as the Fourier transform of the autocorrelation function,
as stated by the Wiener-Khinchin theorem).

When for each frequency bin we average out the power for a given set of 
GRBs, we implicitly assume that each time profile is an individual
realisation of a common stochastic process. On the contrary, if one
is interested in studying the light curve of a single GRB, and treats it
like a deterministic signal affected by uncorrelated noise, the
averaging process does not make sense any more \citep[e.g.,][]{Guidorzi11}.

Under the assumption of a unique stochastic process explaining the
variety of observed GRB light curves, we study the power distribution
as a function of frequency. BSS98 found that for peak normalisation
for a given frequency bin $f_j$, the fluctuations $P_i(f_j)$ around
the average power $\bar P(f_j)$ ($i$ running over a given set of GRBs)
are minimal, and the distribution is an exponential,
$dN/dP(f_j)=N\,\exp{[-P(f_j)/\bar P(f_j)]}$.
For each grouped frequency bin we investigated the observed 
distribution. For both normalisations we did not
subtract the Poisson noise.
%
% ++++++++++++++ Figure 8 ++++++++++++++
\begin{figure}
\centering
\includegraphics[height=5cm]{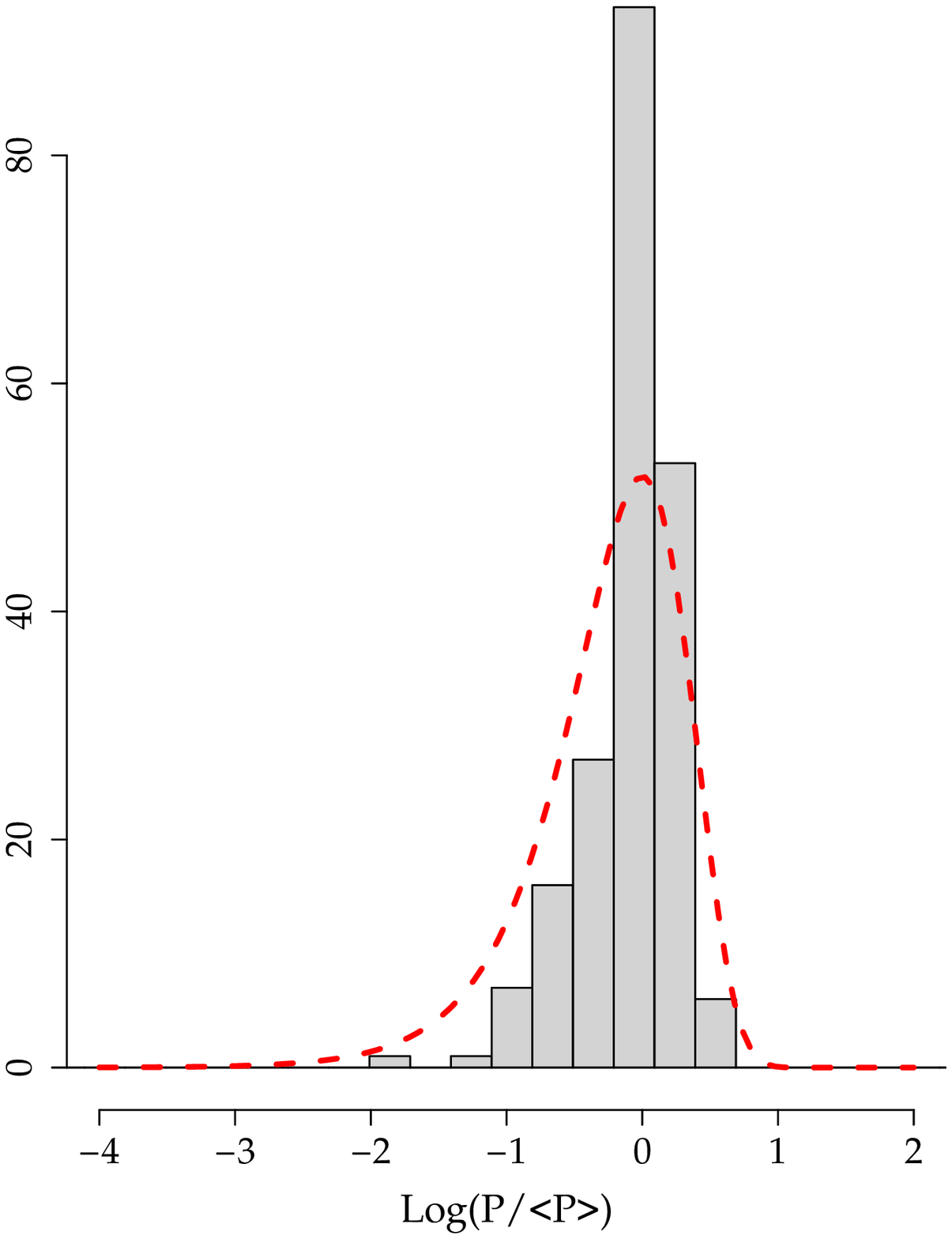}
\includegraphics[height=5cm]{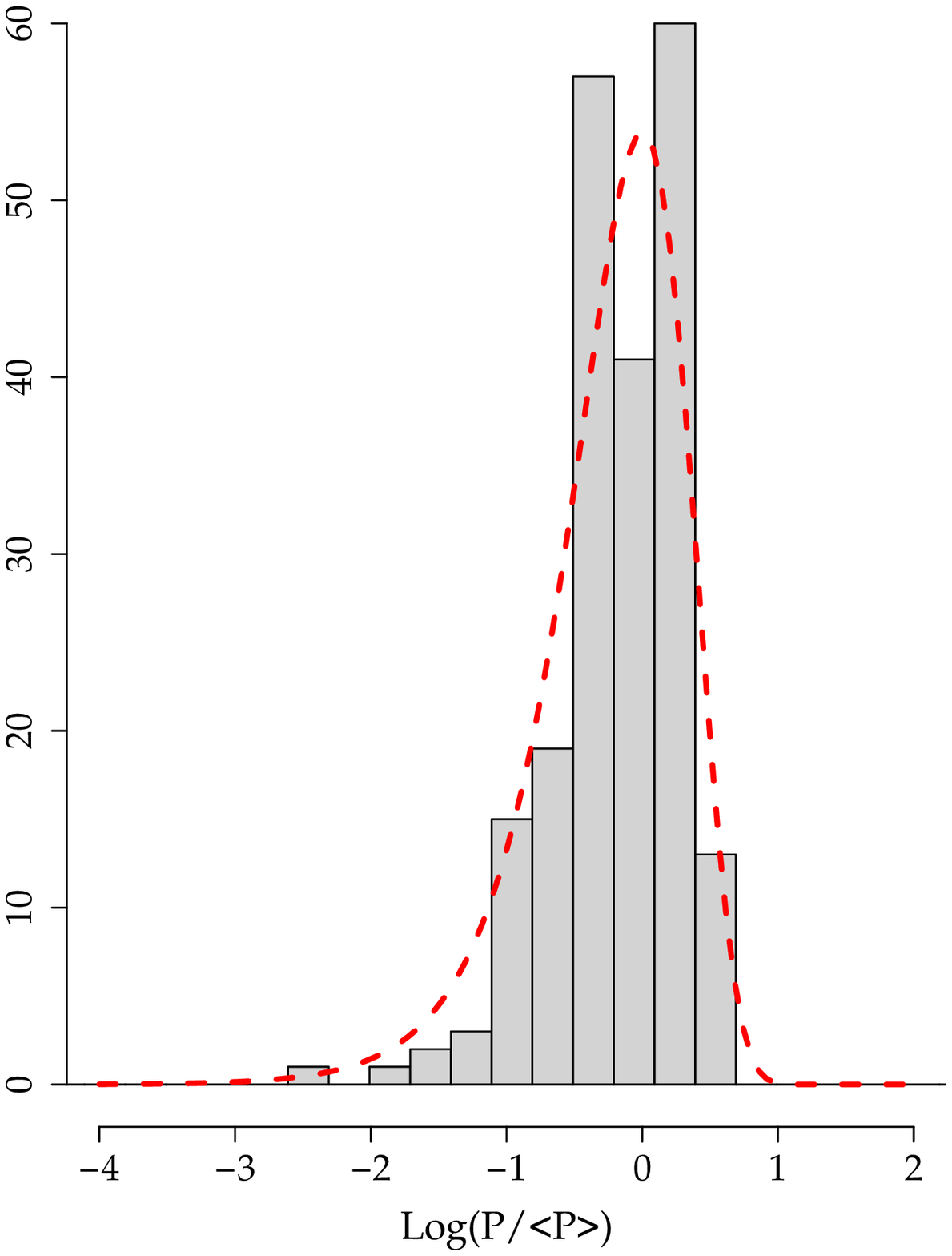}
\includegraphics[height=5cm]{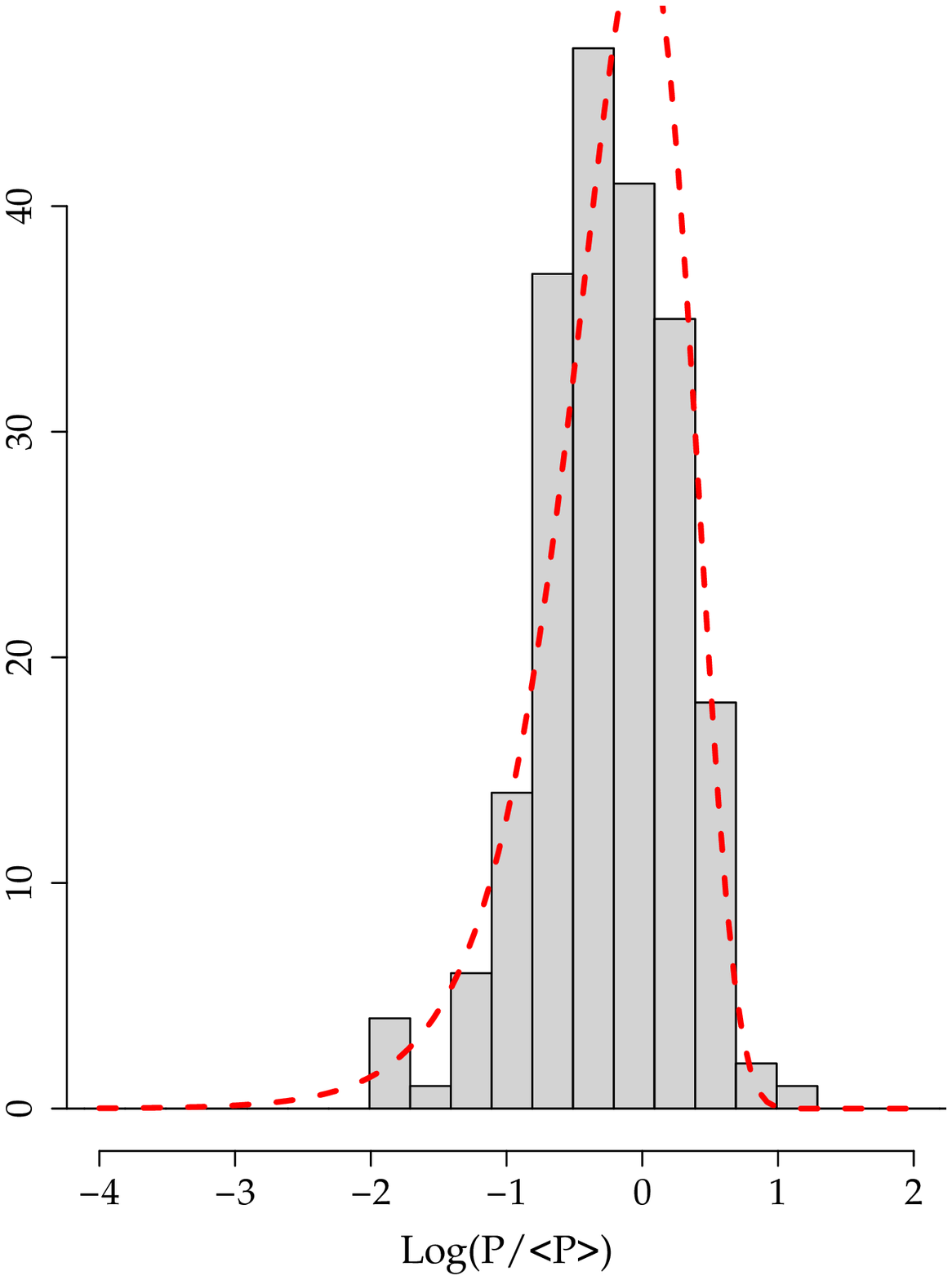}
\includegraphics[height=5cm]{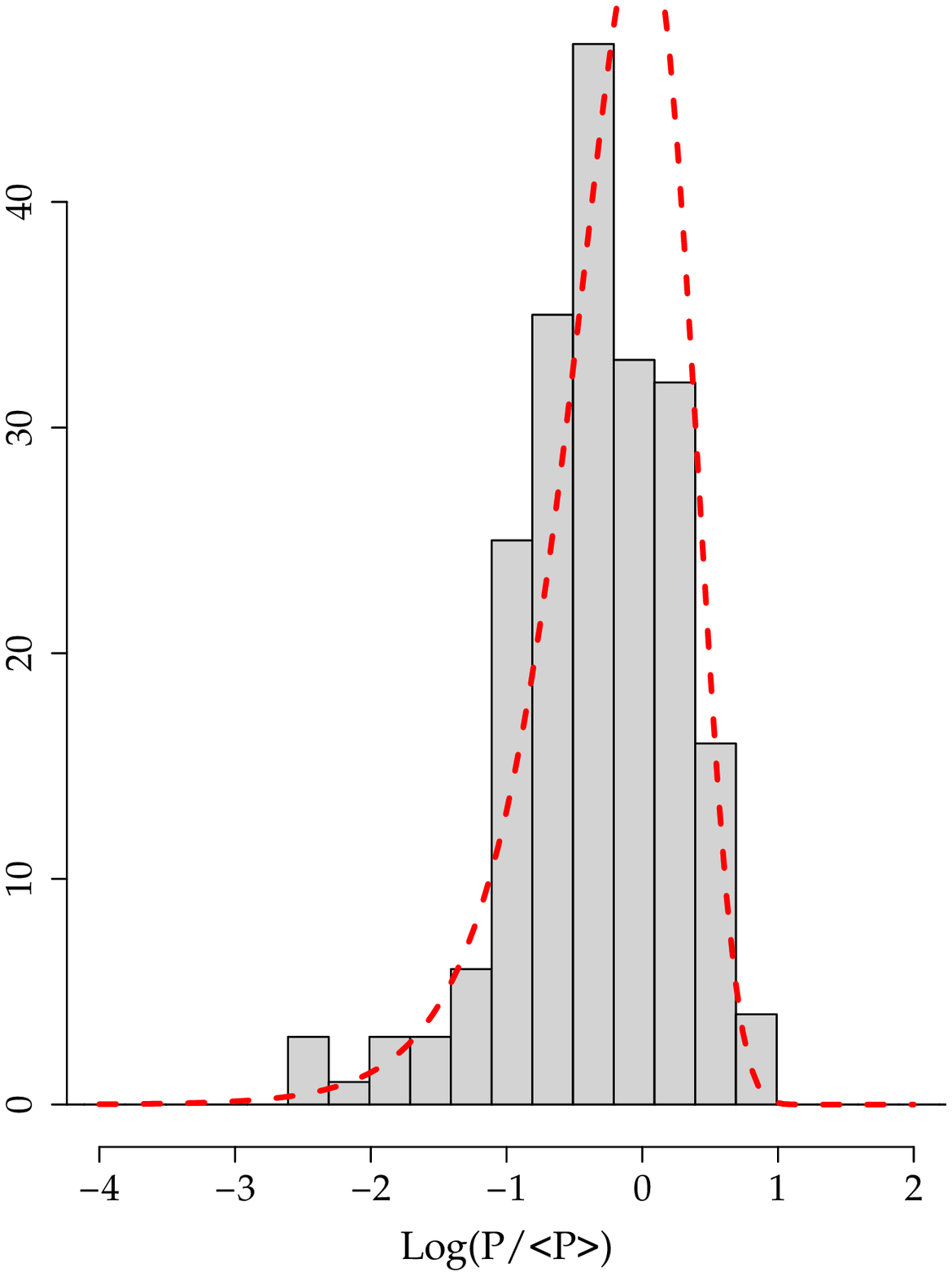}
\caption{Observed distributions of the power at four different
frequency bins from the full sample: from top to bottom, left to right,
$0.01$--$0.02$, $0.05$--$0.06$, $0.23$--$0.24$, and $0.40$--$0.41$~Hz,
respectively. Dashed lines show the corresponding expected exponential
distribution, $dN/dP\propto \exp{(-P/\bar P)}$.}
\label{fig8}
\end{figure}
%++++++++++++++++++++++++++++++++++++++++++
%
Figure~\ref{fig8} displays four distributions corresponding to four different
frequency bins for the full sample of GRBs in the variance normalisation.
We adopted the Anderson--Darling test \citep{Anderson52}
implemented under the $R$ package
{\tt ADGofTest}\footnote{http://cran.r-project.org/web/packages/ADGofTest/.}
(v0.1) to test the compatibility with an exponential.
This test is particularly sensitive to the distribution
tails and, as such, to the possible presence of a few outliers.
The corresponding probability was evaluated as a function of frequency
for both normalisations and for different GRB sets.
The results are shown in the top panels of Figure~\ref{fig9}.
We also studied the ratio between the standard deviation of the observed
distribution and that expected in the exponential distribution hypothesis
(bottom panels of Fig.~\ref{fig9}).
%
% ++++++++++++++ Figure 9 ++++++++++++++
\begin{figure}
\centering
\includegraphics[width=7cm]{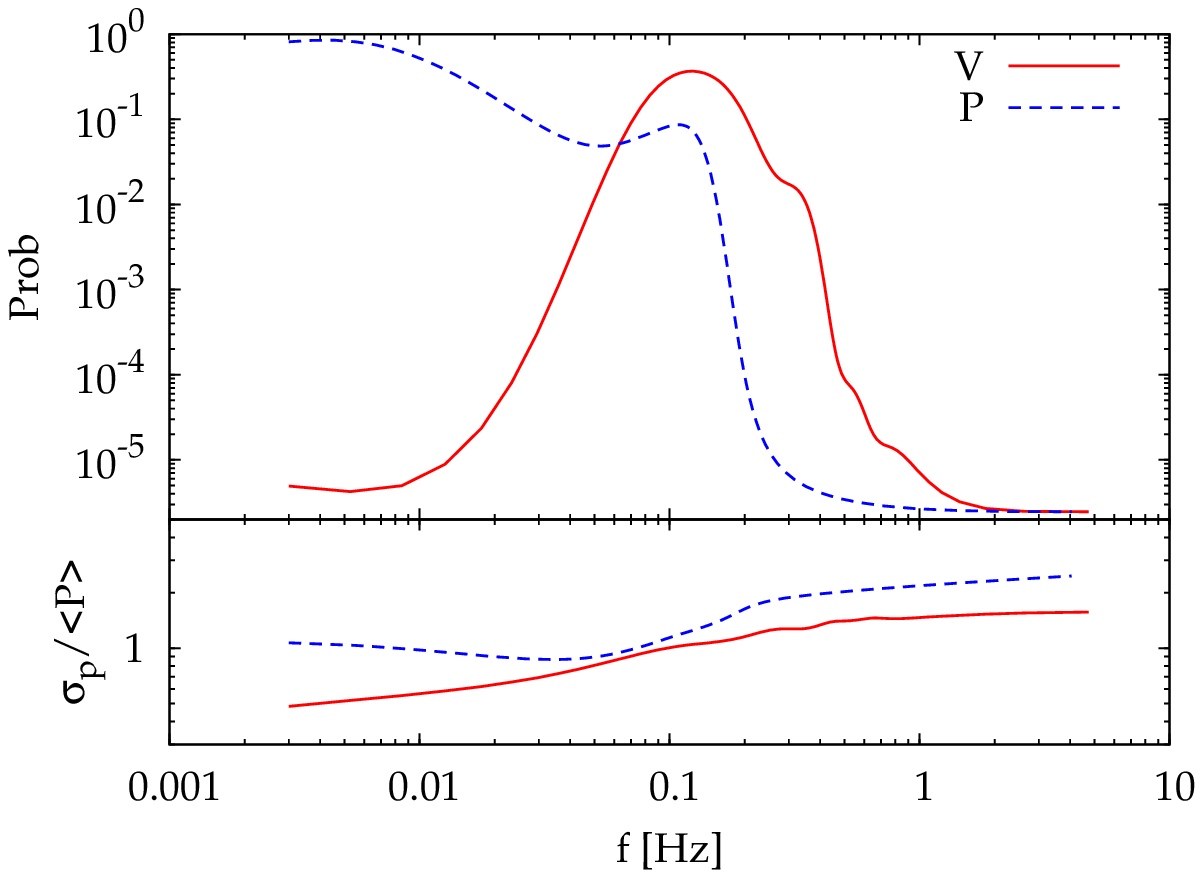}
\includegraphics[width=7cm]{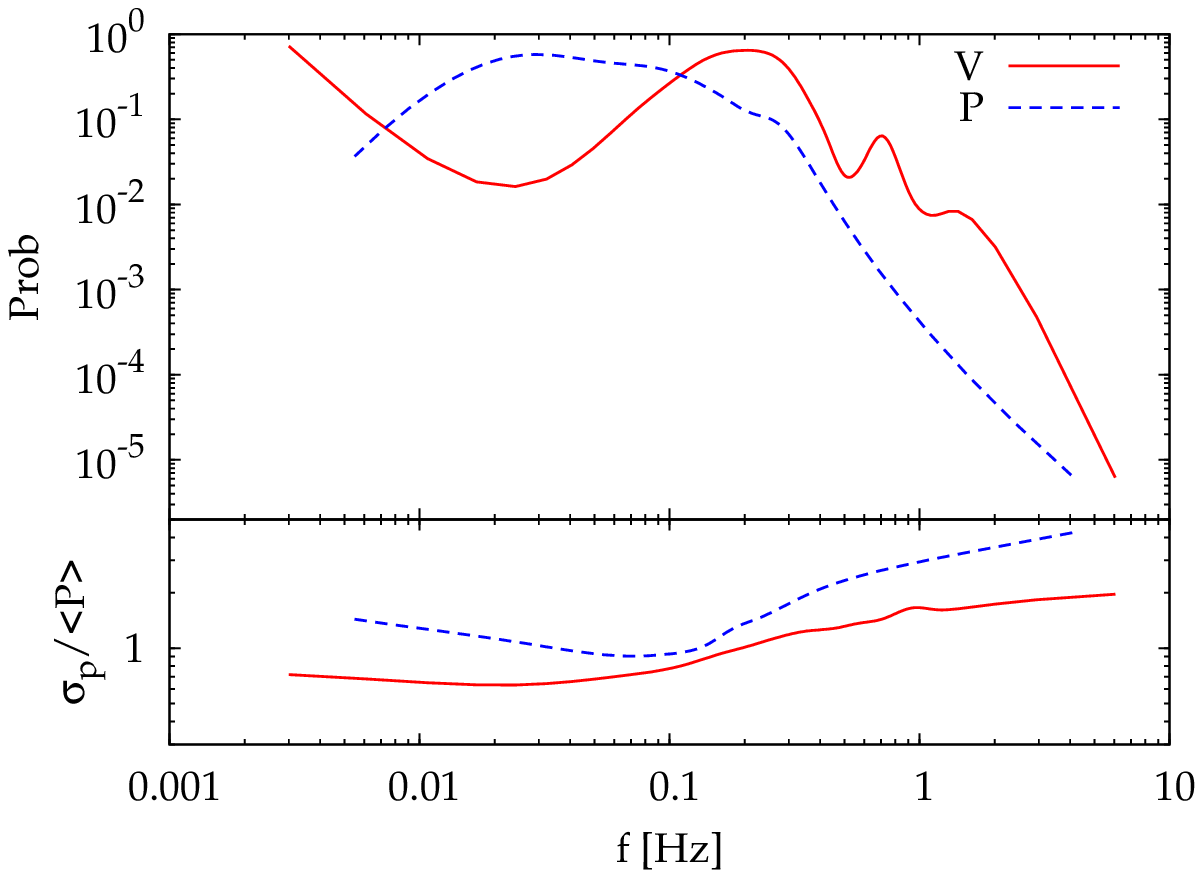}
\includegraphics[width=7cm]{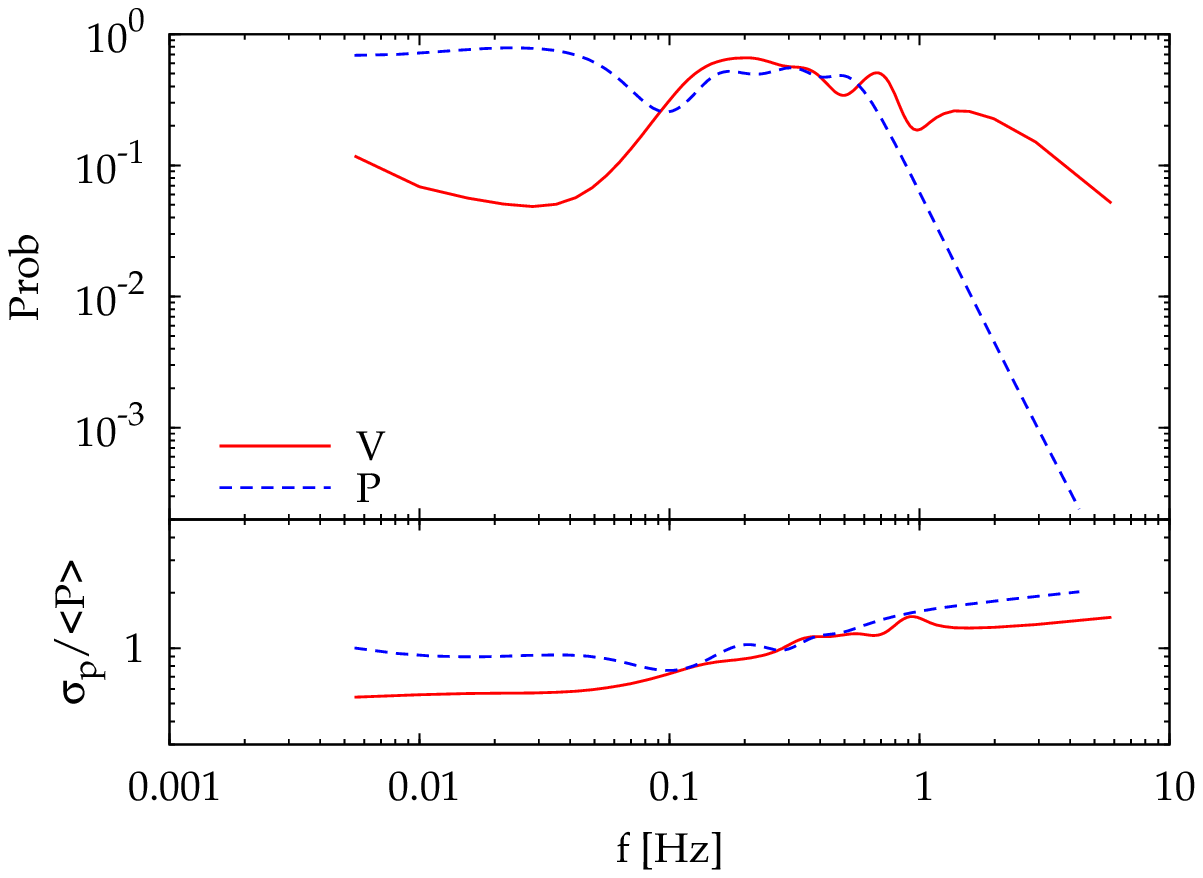}
\caption{{\em Top panel}: P-value based on the Anderson-Darling test that
the corresponding density distribution of power as a function of frequency
is consistent with an exponential. 
The ratio between the $\sigma$ of the observed power distribution
over the corresponding average value is also shown.
The solid (dashed) line shows the variance (peak)
normalisation case. GRBs are from the full sample.
{\em Mid panel}: the same, with GRBs being taken from the $z$-silver sample.
{\em Bottom panel}: the same, with GRBs from the $z$-golden sample.}
\label{fig9}
\end{figure}
%++++++++++++++++++++++++++++++++++++++++++
%

For all the GRB sets the fluctuations of the variance normalisation
are systematically lower than in the peak normalisation.
In the full sample case (top panel of Fig.~\ref{fig9}), at low frequencies
the p-value of the variance normalisation is $\approx10^{-5}$, due to the
small dispersion of the distribution compared to that expected for an
exponential. At $f\ga$~a few~$10^{-2}$~Hz the p-value rises above
$0.01$, and finally decreases again below $0.01$ at $f>$~a few~$0.1$~Hz.
For less numerous sets, such as the $z$-silver and golden samples
(mid and bottom panels of Fig.~\ref{fig9}, respectively), the
exponential hypothesis cannot be rejected at almost any frequency
for the variance normalisation. For the peak normalisation, the
high-frequency range ($f\ga 1$~Hz) exhibits systematically
worse p-values and larger fluctuations around the average power.

In conclusion, the variance normalisation shows minimal fluctuations
around the average power and the PDS is distributed most consistently
with an exponential.
This is also a $\chi^2$ distribution with two
degrees of freedom ($\chi^2_2$), and suggests that the properly
normalised power of each GRB at any given frequency is the result
of a Fourier transform term, whose amplitude is normally distributed
around the average value and the phase is independently and uniformly
distributed \citep{Klis89}.
The fluctuations around the mean value become significantly larger
than what expected for a $\chi^2_2$ distribution above $\sim$1~Hz
in the largest samples (Fig.~\ref{fig9}). In our data at these
frequencies the statistical noise becomes comparable with signal,
and this could be connected with it, although the details are not clear.
Alternatively, it could be suggestive of the presence of a fraction of
GRBs with significantly more power at high frequencies than the
bulk of GRBs.

Concerning the relative weight of some GRBs in driving the average results,
none of the classes considered above (Sections~\ref{sec:res_dur},
\ref{sec:res_flu}, and \ref{sec:res_eiso}) seems to be dominant in
determining the observed properties, especially the high-frequency slope $\alpha_2$.
By definition, the variance normalisation equally weighs each GRB in the
average (noise--subtracted) PDS, since the normalised PDSs of both
bright and dim GRBs have the same area. The only difference is that
dim GRBs have a lower S/N, and, as a consequence, they can contribute
more than bright GRBs to the observed scatter in the power distribution.

%%%%%%%%%%%%%%%%%%%%%%%%%%%%%%%%%%%%%%%%%%%%%%%%%%%%%%%%%%%
\section{Conclusions}
\label{sec:con}
%%%%%%%%%%%%%%%%%%%%%%%%%%%%%%%%%%%%%%%%%%%%%%%%%%%%%%%%%%%
For the first time it was possible to study the average PDS of a
sample of long GRBs by correcting for the cosmological time dilation
effects both on timing and spectral properties.
This is described by a smoothed broken power-law, with a typical
low- (high-) frequency index around $1.0$ ($1.7$--$1.8$), and a break
frequency of a few~$\times10^{-2}$~Hz.
This is mainly determined by an average rest--frame characteristic time
of 2--4~s for the individual shot most GRBs are made of.
We found no clear difference from what is obtained by processing
the same sample within the observer frame, apart from the break
frequency $f_{\rm b}$. In particular, for a restricted sample of
64 GRBs with known $E_{\rm iso}$, $L_{\rm p,iso}$ we found no correlation
between intrinsic properties and the average rest--frame PDS.

Comparing different energy bands, in agreement with previous results
we found that the average PDS in the harder energy channel exhibits
shallower indices, especially at high frequencies.
From the sample of GRBs with known redshift, we found that
in the observed 15--150~keV energy band the average PDS of the high-$z$
GRBs, which also have lower fluences on average, exhibits marginally
shallower indices than the average PDS of the low-$z$/higher-fluence
GRBs. In principle, this can be explained by the farthest GRBs
being observed in source-rest frame harder energy bands.
Whether this property is entirely due to the cosmological energy band
shift bias, or it implies some evolution in the average PDS
with redshift, cannot be settled with the present dataset, because
of the narrow passband of BAT.

In the observer frame the shortest GRBs ($3$~s~$<T_{90}<40$~s), which
on average have lower fluences as well, are characterised by higher
values of $f_{\rm b}$ and shallower values of the low-frequency
power-law index.
This suggests that, within the covered frequency range, they better
approach the asymptotic value of a flat PDS at low frequencies.

In most cases the high-frequency index for different GRB samples
is compatible with $5/3$ expected for a Kolmogorov velocity spectrum
in a turbulent medium, but in fewer cases can also be as steep
as $\sim2$ with no break in the power-law up to several Hz.

At variance with past results, we do not find evidence for any
cut-off around 1--2~Hz in the average PDS. Instead, for the full
and the $z$-silver samples the average PDS is consistent with an
unbroken power-law up to several~Hz at the least (Figs.~\ref{fig2}
and \ref{fig3}). However, the softer energy band of BAT compared
with that of BATSE might account for the missing cut-off above 1~Hz.

Theoretical interpretations of the power-law PDS with an index
compatible with $5/3$ have been proposed in several different
contexts. Within the internal shock model, by tuning the flow of mass and
energy emission through the wind of shells it is possible to
obtain the average power-law in the PDS \citep{Panaitescu99,Spada00}.
The observed PDS index is also obtained from emission due to a
relativistic outflow of a jet propagating through the stellar
and the circumstellar matter \citep{Zhang09,Morsony10}.

Alternatively to the classical internal shock model in which the energy
of the ejecta is mostly kinetic, in the magnetically-dominated outflows
the energy dissipation via magnetic reconnection plays a crucial role
(e.g., Lyutikov 2006; Zhang \& Yan 2011; and references therein).
\nocite{Lyutikov06,ICMART}
In the ICMART model, the rapid reconfiguration of the magnetic field
can trigger MHD turbulence, which ends up in a runaway release
of synchrotron gamma-rays radiated by the accelerated particles.
Unlike the hydrodynamical turbulence characterised
by the $5/3$ Kolmogorov velocity spectrum, MHD turbulence scales
differently for different directions with respect to the field lines:
the index ranges from $5/3$ to $2$ moving from perpendicular
to parallel direction \citep{ICMART}. Interestingly, this range
matches our results.

Within some models, the observed variability may track
that of the progenitor, e.g. through erratic accretion episodes
(e.g., Kumar et al. 2008)\nocite{Kumar08}, or hydrodynamical
or magnetic instabilities in the accretion disc
(e.g., Perna et al. 2006; Proga \& Zhang 2006; Margutti et al. 2011).
\nocite{Perna06,Proga06,Margutti11}
In particular, variability can also arise from turbulence within
the accretion disc. For instance, in the context of magneto-rotational
instability, \citet{Carballido11} studied how neutrino cooling
can shape different PDSs of the observed luminosity, depending
on the cooling process.
Neutrino emissivity scales with temperature $T$ as
$\dot{q}\propto T^\beta$, where $\beta$ is 9 (6) when
$e^\pm$ pair annihilation ($e^\pm$ capture by free neutrons or protons)
is the dominant process. They found that the range of power-law
indices expected in the average PDS from $0.1$ to $100$~Hz vary
between $1.7$ and $2.0$ with no clear break around $1$--$2$~Hz,
and with some bumps above $1$--$10$~Hz which flatten the decline.
Most of our values for the high-frequency power-law index are
closer to $1.7$ (Table~\ref{tbl-4}), and according to the
neutrino-cooling interpretation, on average this would favour larger
values for $\beta$, i.e. where pair annihilation is dominant.
However, due to S/N limitations, our present data set do not allow
us to explore the average properties at high ($>10$~Hz) frequencies,
and the GRBs with sufficient signal are too few to draw statistically
sound conclusions.
Hopefully, in the future more numerous samples of GRBs with high S/N
especially at high ($>10$~Hz) frequencies will allow us to better
discriminate between competing models.

\section*{Acknowledgments}
C.G. acknowledges ASI for financial support (ASI-INAF contract I/088/06/0).

%%%%%%%%%%%%%%%%%%%%%%%%%%%%%%%%%
% REFERENCES
%%%%%%%%%%%%%%%%%%%%%%%%%%%%%%%%%

%%%%%%%%%%% full sample (244 GRBs) %%%%%%%%%%%%%%%%%%%
\onecolumn
\begin{deluxetable}{lrrrrr}
\tablecolumns{6}
\tabletypesize{\scriptsize}
%\rotate
\tablecaption{Table of the full sample including 244 GRBs in the observer frame.
\label{tbl-1}}
\tablewidth{0pt}
\tablehead{
\colhead{GRB} & \colhead{$t_{\rm start}$\tablenotemark{a}} & \colhead{$t_{\rm stop}$\tablenotemark{a}} & \colhead{Log(Count Fluence)} & \colhead{Log(peak rate)} & \colhead{$T_{90}$\tablenotemark{a}}\\
 & (s) & (s) & \colhead{(count~det$^{-1}$)} & \colhead{(count~s$^{-1}$~det$^{-1}$)} & \colhead{(s)}}
\startdata
050117   & $-203.8$ & $ 302.7$ & $   1.155\pm   0.007$ & $  -0.486\pm   0.029$ & $ 200.0$\\
050124   & $  -3.7$ & $   6.6$ & $   0.265\pm   0.015$ & $  -0.098\pm   0.033$ & $   4.1$\\
050128   & $ -37.3$ & $  52.8$ & $   0.851\pm   0.015$ & $   0.158\pm   0.042$ & $  13.8$\\
050219A  & $ -30.3$ & $  47.0$ & $   0.772\pm   0.011$ & $  -0.357\pm   0.028$ & $  23.0$\\
050219B  & $ -96.1$ & $ 106.8$ & $   1.354\pm   0.010$ & $   0.584\pm   0.028$ & $  27.0$\\
050306   & $-188.1$ & $ 302.8$ & $   1.212\pm   0.015$ & $  -0.362\pm   0.041$ & $ 158.4$\\
050315   & $-179.8$ & $ 191.6$ & $   0.795\pm   0.015$ & $  -0.673\pm   0.037$ & $  40.0$\\
050319   & $-160.0$ & $ 187.0$ & $   0.406\pm   0.030$ & $  -0.767\pm   0.043$ & $  10.0$\\
050326   & $ -49.2$ & $  68.9$ & $   1.120\pm   0.005$ & $   0.284\pm   0.018$ & $  29.5$\\
050401   & $ -43.5$ & $  64.8$ & $   1.105\pm   0.014$ & $   0.176\pm   0.030$ & $  33.0$\\
050418   & $ -92.8$ & $ 163.9$ & $   0.936\pm   0.012$ & $  -0.339\pm   0.022$ & $  83.0$\\
050502B  & $ -34.6$ & $  20.7$ & $  -0.085\pm   0.031$ & $  -0.616\pm   0.034$ & $  17.5$\\
050505   & $ -63.8$ & $  97.3$ & $   0.550\pm   0.028$ & $  -0.684\pm   0.042$ & $  60.0$\\
050525A  & $ -12.3$ & $  25.1$ & $   1.416\pm   0.002$ & $   0.829\pm   0.007$ & $   5.2$\\
050603   & $ -26.4$ & $  41.7$ & $   1.078\pm   0.013$ & $   0.846\pm   0.027$ & $  12.0$\\
050607   & $ -19.0$ & $  37.3$ & $   0.057\pm   0.027$ & $  -0.902\pm   0.041$ & $  26.5$\\
050701   & $ -10.1$ & $  33.0$ & $   0.411\pm   0.011$ & $  -0.420\pm   0.026$ & $  40.0$\\
050713A  & $-112.6$ & $ 223.9$ & $   0.911\pm   0.012$ & $  -0.211\pm   0.024$ & $  70.0$\\
050713B  & $-109.7$ & $ 230.4$ & $   0.875\pm   0.032$ & $  -0.717\pm   0.050$ & $  75.0$\\
050715   & $-150.0$ & $ 224.0$ & $   0.529\pm   0.024$ & $  -0.843\pm   0.039$ & $  52.0$\\
050716   & $ -93.8$ & $ 147.1$ & $   0.979\pm   0.013$ & $  -0.608\pm   0.030$ & $  69.0$\\
050717   & $ -89.0$ & $ 178.1$ & $   0.973\pm   0.008$ & $   0.065\pm   0.038$ & $  86.0$\\
050801   & $  -6.2$ & $  12.8$ & $  -0.289\pm   0.034$ & $  -0.675\pm   0.039$ & $  20.0$\\
050803   & $ -24.2$ & $ 234.5$ & $   0.573\pm   0.022$ & $  -0.790\pm   0.045$ & $  85.0$\\
050820B  & $ -14.6$ & $  29.9$ & $   0.528\pm   0.008$ & $  -0.271\pm   0.021$ & $  13.0$\\
050822   & $ -65.0$ & $ 206.8$ & $   0.732\pm   0.022$ & $  -0.593\pm   0.037$ & $ 102.0$\\
050827   & $ -69.3$ & $  76.1$ & $   0.530\pm   0.019$ & $  -0.617\pm   0.031$ & $  49.0$\\
050911   & $ -17.2$ & $  32.1$ & $  -0.273\pm   0.073$ & $  -0.538\pm   0.055$ & $  16.0$\\
050915B  & $ -50.0$ & $  98.4$ & $   0.822\pm   0.010$ & $  -0.537\pm   0.028$ & $  40.0$\\
050922C  & $  -8.1$ & $   9.1$ & $   0.411\pm   0.010$ & $   0.050\pm   0.025$ & $   5.0$\\
051006   & $ -28.7$ & $  49.8$ & $   0.396\pm   0.029$ & $  -0.346\pm   0.065$ & $  26.0$\\
051111   & $ -60.7$ & $ 103.1$ & $   0.797\pm   0.011$ & $  -0.377\pm   0.029$ & $  47.0$\\
051113   & $-130.2$ & $ 165.1$ & $   0.664\pm   0.030$ & $  -0.536\pm   0.046$ & $  94.0$\\
051227   & $ -34.2$ & $  68.7$ & $  -0.061\pm   0.044$ & $  -0.839\pm   0.051$ & $   8.0$\\
060105   & $-100.0$ & $ 153.4$ & $   1.393\pm   0.004$ & $   0.083\pm   0.028$ & $  55.0$\\
060110   & $ -27.4$ & $  50.4$ & $   0.462\pm   0.013$ & $  -0.560\pm   0.024$ & $  17.0$\\
060111B  & $ -61.6$ & $ 117.3$ & $   0.379\pm   0.034$ & $  -0.726\pm   0.048$ & $  59.0$\\
060115   & $-118.1$ & $ 224.4$ & $   0.546\pm   0.021$ & $  -0.993\pm   0.034$ & $ 142.0$\\
060117   & $ -28.8$ & $  52.1$ & $   1.502\pm   0.003$ & $   0.867\pm   0.017$ & $  16.0$\\
060204B  & $-166.5$ & $ 264.6$ & $   0.703\pm   0.017$ & $  -0.758\pm   0.032$ & $ 134.0$\\
060206   & $ -11.4$ & $  21.8$ & $   0.211\pm   0.013$ & $  -0.406\pm   0.018$ & $   7.0$\\
060210   & $-299.8$ & $ 302.3$ & $   1.093\pm   0.015$ & $  -0.349\pm   0.041$ & $  46.0$\\
060223A  & $ -14.1$ & $  22.2$ & $   0.058\pm   0.027$ & $  -0.782\pm   0.039$ & $  11.0$\\
060223B  & $ -17.7$ & $  16.6$ & $   0.417\pm   0.012$ & $  -0.280\pm   0.038$ & $  10.2$\\
060306   & $ -63.8$ & $ 124.1$ & $   0.560\pm   0.020$ & $   0.022\pm   0.028$ & $  61.0$\\
060312   & $-100.5$ & $ 118.2$ & $   0.597\pm   0.015$ & $  -0.706\pm   0.029$ & $  43.0$\\
060322   & $-239.2$ & $ 477.8$ & $   0.915\pm   0.017$ & $  -0.606\pm   0.025$ & $ 213.0$\\
060403   & $ -37.0$ & $  55.5$ & $   0.306\pm   0.020$ & $  -0.904\pm   0.031$ & $  30.0$\\
060413   & $ -60.0$ & $ 320.0$ & $   0.846\pm   0.011$ & $  -0.900\pm   0.024$ & $ 150.0$\\
060418   & $-171.4$ & $ 285.8$ & $   1.160\pm   0.009$ & $   0.073\pm   0.038$ & $  52.0$\\
060421   & $ -16.4$ & $  24.5$ & $   0.287\pm   0.015$ & $  -0.397\pm   0.027$ & $  11.0$\\
060424   & $ -58.0$ & $  59.4$ & $   0.109\pm   0.042$ & $  -0.521\pm   0.049$ & $  37.0$\\
060428A  & $ -54.8$ & $  89.9$ & $   0.456\pm   0.017$ & $  -0.571\pm   0.026$ & $  39.4$\\
060502A  & $ -29.1$ & $  41.5$ & $   0.555\pm   0.013$ & $  -0.685\pm   0.021$ & $  33.0$\\
060507   & $-202.8$ & $ 380.3$ & $   0.916\pm   0.017$ & $  -0.840\pm   0.036$ & $ 185.0$\\
060510A  & $ -23.3$ & $  29.5$ & $   1.229\pm   0.018$ & $   0.422\pm   0.037$ & $  21.0$\\
060526   & $-239.2$ & $ 518.0$ & $   0.432\pm   0.042$ & $  -0.639\pm   0.043$ & $  13.8$\\
060607A  & $-126.1$ & $ 214.6$ & $   0.665\pm   0.015$ & $  -0.752\pm   0.020$ & $ 100.0$\\
060607B  & $ -30.5$ & $  62.0$ & $   0.425\pm   0.025$ & $  -0.807\pm   0.050$ & $  31.0$\\
060614   & $-166.2$ & $ 327.7$ & $   1.599\pm   0.004$ & $   0.441\pm   0.036$ & $ 102.0$\\
060707   & $-114.9$ & $  86.3$ & $   0.496\pm   0.027$ & $  -0.930\pm   0.041$ & $  68.0$\\
060708   & $  -5.8$ & $  11.8$ & $  -0.049\pm   0.018$ & $  -0.573\pm   0.034$ & $   9.8$\\
060714   & $-134.2$ & $ 233.6$ & $   0.738\pm   0.021$ & $  -0.746\pm   0.039$ & $ 115.0$\\
060719   & $ -59.9$ & $ 119.6$ & $   0.431\pm   0.020$ & $  -0.529\pm   0.034$ & $  55.0$\\
060813   & $ -28.2$ & $  55.7$ & $   0.923\pm   0.006$ & $   0.054\pm   0.012$ & $  14.9$\\
060814   & $ -74.9$ & $ 384.9$ & $   1.386\pm   0.004$ & $  -0.031\pm   0.015$ & $ 146.0$\\
060825   & $ -10.2$ & $  14.5$ & $   0.269\pm   0.012$ & $  -0.438\pm   0.026$ & $   8.1$\\
060904A  & $ -60.0$ & $ 208.1$ & $   1.133\pm   0.005$ & $  -0.130\pm   0.023$ & $  80.0$\\
060904B  & $-173.7$ & $ 347.0$ & $   0.453\pm   0.036$ & $  -0.467\pm   0.031$ & $ 192.0$\\
060908   & $ -29.3$ & $  32.2$ & $   0.606\pm   0.012$ & $  -0.305\pm   0.033$ & $  19.3$\\
060912A  & $  -5.5$ & $  10.3$ & $   0.381\pm   0.013$ & $   0.049\pm   0.021$ & $   5.0$\\
060927   & $ -24.7$ & $  47.0$ & $   0.325\pm   0.017$ & $  -0.404\pm   0.023$ & $  22.6$\\
061004   & $  -5.1$ & $  12.5$ & $   0.032\pm   0.016$ & $  -0.399\pm   0.027$ & $   6.2$\\
061007   & $ -39.9$ & $ 357.9$ & $   1.802\pm   0.002$ & $   0.431\pm   0.023$ & $  75.0$\\
061019   & $-239.5$ & $  64.0$ & $   0.683\pm   0.058$ & $  -0.700\pm   0.056$ & $ 191.0$\\
061021   & $ -54.5$ & $ 109.2$ & $   0.673\pm   0.012$ & $  -0.059\pm   0.018$ & $  46.0$\\
061121   & $-135.1$ & $ 268.2$ & $   1.372\pm   0.003$ & $   0.546\pm   0.014$ & $  81.0$\\
061126   & $ -86.2$ & $ 156.8$ & $   0.998\pm   0.009$ & $   0.147\pm   0.017$ & $ 191.0$\\
061202   & $   9.8$ & $ 196.2$ & $   0.574\pm   0.014$ & $  -0.656\pm   0.028$ & $  91.0$\\
061222A  & $-113.5$ & $ 219.6$ & $   1.131\pm   0.006$ & $   0.270\pm   0.019$ & $  72.0$\\
070103   & $ -18.2$ & $  37.1$ & $  -0.191\pm   0.047$ & $  -0.676\pm   0.064$ & $  19.0$\\
070107   & $ -24.0$ & $ 662.2$ & $   0.902\pm   0.015$ & $  -0.637\pm   0.026$ & $ 333.8$\\
070220   & $-187.1$ & $ 345.9$ & $   1.216\pm   0.007$ & $  -0.098\pm   0.020$ & $ 129.0$\\
070306   & $-239.2$ & $ 487.7$ & $   0.991\pm   0.015$ & $  -0.295\pm   0.019$ & $ 210.0$\\
070318   & $ -63.9$ & $ 125.6$ & $   0.628\pm   0.013$ & $  -0.624\pm   0.024$ & $  63.0$\\
070328   & $ -90.0$ & $ 227.2$ & $   1.126\pm   0.006$ & $  -0.256\pm   0.020$ & $  69.0$\\
070411   & $-124.1$ & $ 197.7$ & $   0.704\pm   0.016$ & $  -0.948\pm   0.035$ & $ 101.0$\\
070419B  & $-239.9$ & $ 400.0$ & $   1.118\pm   0.008$ & $  -0.765\pm   0.023$ & $ 236.5$\\
070420   & $-187.4$ & $ 230.4$ & $   1.360\pm   0.010$ & $  -0.093\pm   0.026$ & $  77.0$\\
070427   & $ -11.0$ & $  23.8$ & $   0.159\pm   0.015$ & $  -0.716\pm   0.038$ & $  11.0$\\
070508   & $ -46.6$ & $  73.1$ & $   1.464\pm   0.002$ & $   0.572\pm   0.015$ & $  20.9$\\
070521   & $ -70.3$ & $  97.7$ & $   1.096\pm   0.006$ & $   0.060\pm   0.022$ & $  37.9$\\
070529   & $-115.9$ & $ 237.8$ & $   0.638\pm   0.033$ & $  -0.659\pm   0.071$ & $ 109.0$\\
070612A  & $-239.0$ & $ 514.2$ & $   1.227\pm   0.018$ & $  -0.805\pm   0.030$ & $ 370.0$\\
070616   & $ -89.9$ & $ 963.1$ & $   1.528\pm   0.004$ & $  -0.563\pm   0.029$ & $ 402.0$\\
070621   & $ -45.5$ & $  77.0$ & $   0.845\pm   0.011$ & $  -0.449\pm   0.038$ & $  33.3$\\
070628   & $ -76.2$ & $  59.0$ & $   0.765\pm   0.016$ & $  -0.264\pm   0.017$ & $  39.1$\\
070704   & $-239.0$ & $ 732.8$ & $   0.999\pm   0.013$ & $  -0.592\pm   0.017$ & $ 380.0$\\
070721B  & $-239.4$ & $ 542.7$ & $   0.757\pm   0.025$ & $  -0.720\pm   0.036$ & $ 340.0$\\
070808   & $ -28.9$ & $  55.2$ & $   0.265\pm   0.027$ & $  -0.573\pm   0.036$ & $  32.0$\\
070911   & $-283.5$ & $ 302.8$ & $   1.347\pm   0.005$ & $  -0.210\pm   0.026$ & $ 162.0$\\
070917   & $ -11.4$ & $  22.6$ & $   0.489\pm   0.009$ & $   0.054\pm   0.016$ & $   7.3$\\
071001   & $ -54.6$ & $ 110.7$ & $   0.208\pm   0.031$ & $  -0.985\pm   0.046$ & $  58.5$\\
071003   & $-182.6$ & $ 342.7$ & $   1.036\pm   0.014$ & $  -0.139\pm   0.024$ & $ 150.0$\\
071010B  & $ -75.6$ & $  77.1$ & $   0.861\pm   0.007$ & $  -0.111\pm   0.012$ & $  35.7$\\
071011   & $ -64.7$ & $ 119.3$ & $   0.495\pm   0.029$ & $  -0.714\pm   0.056$ & $  61.0$\\
071020   & $ -12.8$ & $  16.9$ & $   0.507\pm   0.007$ & $   0.119\pm   0.018$ & $   4.2$\\
071025   & $-131.5$ & $ 383.7$ & $   1.067\pm   0.009$ & $  -0.773\pm   0.028$ & $ 109.0$\\
071117   & $  -5.0$ & $   9.4$ & $   0.414\pm   0.010$ & $   0.057\pm   0.023$ & $   6.6$\\
080205   & $-121.4$ & $ 217.1$ & $   0.523\pm   0.027$ & $  -0.899\pm   0.052$ & $ 106.5$\\
080210   & $ -35.0$ & $  39.8$ & $   0.423\pm   0.018$ & $  -0.831\pm   0.029$ & $  45.0$\\
080229A  & $ -69.7$ & $ 132.4$ & $   1.068\pm   0.010$ & $   0.177\pm   0.039$ & $  64.0$\\
080310   & $ -84.0$ & $ 659.5$ & $   0.676\pm   0.025$ & $  -0.953\pm   0.047$ & $ 365.0$\\
080319B  & $ -69.6$ & $  62.2$ & $   2.057\pm   0.001$ & $   0.792\pm   0.018$ & $ 125.0$\\
080319C  & $ -14.3$ & $  27.3$ & $   0.591\pm   0.013$ & $  -0.186\pm   0.034$ & $  34.0$\\
080328   & $-115.9$ & $ 225.1$ & $   1.036\pm   0.010$ & $  -0.277\pm   0.041$ & $  90.6$\\
080409   & $ -10.0$ & $  20.4$ & $  -0.205\pm   0.054$ & $  -0.481\pm   0.043$ & $  20.2$\\
080411   & $ -78.3$ & $ 155.9$ & $   1.647\pm   0.002$ & $   0.743\pm   0.007$ & $  56.0$\\
080413A  & $ -48.2$ & $  97.4$ & $   0.734\pm   0.011$ & $  -0.168\pm   0.015$ & $  46.0$\\
080413B  & $  -6.9$ & $  11.0$ & $   0.627\pm   0.011$ & $   0.280\pm   0.021$ & $   8.0$\\
080430   & $ -20.4$ & $  40.1$ & $   0.360\pm   0.015$ & $  -0.472\pm   0.022$ & $  16.2$\\
080503   & $ -90.8$ & $ 181.6$ & $   0.487\pm   0.023$ & $  -0.564\pm   0.063$ & $ 170.0$\\
080602   & $ -91.8$ & $ 149.3$ & $   0.687\pm   0.029$ & $  -0.326\pm   0.045$ & $  74.0$\\
080603B  & $ -67.5$ & $ 135.4$ & $   0.639\pm   0.012$ & $  -0.226\pm   0.028$ & $  60.0$\\
080605   & $ -39.4$ & $  63.1$ & $   1.297\pm   0.004$ & $   0.505\pm   0.023$ & $  20.0$\\
080607   & $-157.2$ & $ 292.3$ & $   1.556\pm   0.006$ & $   0.577\pm   0.028$ & $  79.0$\\
080613B  & $ -86.4$ & $ 157.2$ & $   0.946\pm   0.007$ & $  -0.224\pm   0.044$ & $ 105.0$\\
080623   & $  -8.7$ & $  16.0$ & $   0.181\pm   0.027$ & $  -0.488\pm   0.059$ & $  15.2$\\
080714   & $ -30.9$ & $  51.2$ & $   0.608\pm   0.014$ & $  -0.252\pm   0.024$ & $  33.0$\\
080721   & $ -34.6$ & $  39.2$ & $   1.264\pm   0.013$ & $   0.513\pm   0.041$ & $  16.2$\\
080725   & $ -50.9$ & $  55.9$ & $   0.715\pm   0.013$ & $  -0.596\pm   0.022$ & $ 120.0$\\
080727B  & $  -9.6$ & $  18.5$ & $   0.648\pm   0.007$ & $   0.114\pm   0.022$ & $   8.6$\\
080727C  & $-113.9$ & $ 222.3$ & $   0.924\pm   0.010$ & $  -0.519\pm   0.024$ & $  79.7$\\
080804   & $ -33.5$ & $  68.6$ & $   0.713\pm   0.018$ & $  -0.444\pm   0.033$ & $  34.0$\\
080805   & $ -88.7$ & $ 170.3$ & $   0.644\pm   0.014$ & $  -0.846\pm   0.027$ & $  78.0$\\
080810   & $-122.6$ & $ 220.2$ & $   0.860\pm   0.012$ & $  -0.499\pm   0.034$ & $ 106.0$\\
080903   & $ -61.9$ & $  98.0$ & $   0.489\pm   0.017$ & $  -0.991\pm   0.055$ & $  66.0$\\
080905B  & $ -96.1$ & $ 191.9$ & $   0.483\pm   0.032$ & $  -0.759\pm   0.045$ & $ 128.0$\\
080906   & $-239.4$ & $ 267.3$ & $   0.830\pm   0.015$ & $  -0.932\pm   0.022$ & $ 147.0$\\
080915B  & $  -4.0$ & $   6.4$ & $   0.259\pm   0.012$ & $   0.025\pm   0.019$ & $   3.9$\\
080916A  & $ -76.4$ & $ 146.2$ & $   0.839\pm   0.008$ & $  -0.461\pm   0.016$ & $  60.0$\\
080928   & $-131.5$ & $ 400.0$ & $   0.675\pm   0.020$ & $  -0.495\pm   0.031$ & $ 280.0$\\
081008   & $-239.6$ & $ 338.4$ & $   0.876\pm   0.013$ & $  -0.797\pm   0.024$ & $ 185.5$\\
081102   & $ -54.7$ & $  54.2$ & $   0.551\pm   0.023$ & $  -0.756\pm   0.037$ & $  63.0$\\
081109   & $-109.0$ & $ 169.0$ & $   0.790\pm   0.011$ & $  -0.791\pm   0.019$ & $ 190.0$\\
081126   & $ -85.3$ & $ 106.9$ & $   0.723\pm   0.012$ & $  -0.302\pm   0.021$ & $  54.0$\\
081128   & $ -80.0$ & $ 153.5$ & $   0.681\pm   0.015$ & $  -0.787\pm   0.037$ & $ 100.0$\\
081203A  & $-141.7$ & $ 255.4$ & $   1.052\pm   0.008$ & $  -0.318\pm   0.032$ & $ 294.0$\\
081210   & $-109.9$ & $ 287.1$ & $   0.546\pm   0.025$ & $  -0.453\pm   0.028$ & $ 146.0$\\
081221   & $-102.4$ & $ 205.4$ & $   1.509\pm   0.003$ & $   0.375\pm   0.013$ & $  33.5$\\
081222   & $ -37.1$ & $  72.7$ & $   0.922\pm   0.005$ & $   0.019\pm   0.012$ & $  24.0$\\
090102   & $ -12.4$ & $  85.2$ & $   1.013\pm   0.017$ & $   0.022\pm   0.053$ & $  27.0$\\
090113   & $  -9.4$ & $  18.7$ & $   0.112\pm   0.020$ & $  -0.367\pm   0.033$ & $   9.1$\\
090123   & $-120.0$ & $ 178.3$ & $   0.658\pm   0.023$ & $  -0.738\pm   0.045$ & $ 131.0$\\
090129   & $ -26.0$ & $  51.8$ & $   0.630\pm   0.010$ & $  -0.297\pm   0.020$ & $  17.5$\\
090201   & $-118.9$ & $ 210.7$ & $   1.660\pm   0.007$ & $   0.356\pm   0.029$ & $  83.0$\\
090301   & $ -93.6$ & $  85.1$ & $   1.515\pm   0.003$ & $   0.418\pm   0.016$ & $  41.0$\\
090401A  & $ -14.0$ & $ 292.1$ & $   1.245\pm   0.007$ & $   0.189\pm   0.020$ & $ 112.0$\\
090401B  & $-140.5$ & $ 281.0$ & $   1.194\pm   0.006$ & $   0.558\pm   0.012$ & $ 183.0$\\
090404   & $-119.6$ & $ 154.0$ & $   0.808\pm   0.011$ & $  -0.586\pm   0.032$ & $  84.0$\\
090410   & $ -76.0$ & $ 328.2$ & $   0.962\pm   0.009$ & $  -0.438\pm   0.033$ & $ 165.0$\\
090418A  & $ -69.6$ & $ 124.7$ & $   0.854\pm   0.013$ & $  -0.545\pm   0.046$ & $  56.0$\\
090422   & $  -8.1$ & $  16.0$ & $  -0.308\pm   0.051$ & $  -0.411\pm   0.046$ & $   8.5$\\
090423   & $ -10.7$ & $  22.3$ & $   0.105\pm   0.019$ & $  -0.679\pm   0.035$ & $  10.3$\\
090424   & $ -83.2$ & $ 164.4$ & $   1.522\pm   0.005$ & $   0.948\pm   0.013$ & $  48.0$\\
090509   & $-239.3$ & $ 300.0$ & $   0.731\pm   0.038$ & $  -0.678\pm   0.062$ & $ 335.0$\\
090516   & $-231.2$ & $ 447.1$ & $   1.114\pm   0.031$ & $  -0.564\pm   0.057$ & $ 210.0$\\
090518   & $  -5.0$ & $   6.4$ & $  -0.113\pm   0.021$ & $  -0.527\pm   0.037$ & $   6.9$\\
090530   & $ -41.0$ & $  81.8$ & $   0.266\pm   0.034$ & $  -0.260\pm   0.046$ & $  48.0$\\
090531B  & $ -55.6$ & $ 112.0$ & $   0.109\pm   0.040$ & $  -0.574\pm   0.044$ & $  80.0$\\
090618   & $-214.5$ & $ 412.3$ & $   2.244\pm   0.001$ & $   0.657\pm   0.003$ & $ 113.2$\\
090628   & $ -19.0$ & $  15.2$ & $   0.058\pm   0.029$ & $  -0.728\pm   0.046$ & $  20.1$\\
090709A  & $-236.5$ & $ 273.5$ & $   1.576\pm   0.003$ & $   0.030\pm   0.016$ & $  89.0$\\
090715B  & $-239.6$ & $ 550.0$ & $   0.993\pm   0.012$ & $  -0.316\pm   0.012$ & $ 266.0$\\
090812   & $ -99.3$ & $ 183.9$ & $   0.939\pm   0.009$ & $  -0.242\pm   0.028$ & $  66.7$\\
090813   & $  -7.0$ & $  13.9$ & $   0.375\pm   0.021$ & $   0.182\pm   0.039$ & $   7.1$\\
090904A  & $ -96.2$ & $ 321.8$ & $   0.801\pm   0.017$ & $  -0.664\pm   0.022$ & $ 122.0$\\
090904B  & $ -60.0$ & $ 105.4$ & $   1.254\pm   0.013$ & $  -0.382\pm   0.081$ & $  47.0$\\
090912   & $-139.6$ & $ 276.2$ & $   0.889\pm   0.021$ & $  -0.774\pm   0.036$ & $ 144.0$\\
090926B  & $-171.2$ & $ 281.4$ & $   1.029\pm   0.012$ & $  -0.477\pm   0.021$ & $ 109.7$\\
090929B  & $-239.4$ & $ 380.0$ & $   1.025\pm   0.014$ & $  -0.190\pm   0.032$ & $ 360.0$\\
091018   & $  -7.0$ & $  13.4$ & $   0.497\pm   0.008$ & $   0.093\pm   0.012$ & $   4.4$\\
091020   & $ -47.1$ & $  77.5$ & $   0.760\pm   0.012$ & $  -0.295\pm   0.019$ & $  34.6$\\
091026   & $ -56.0$ & $ 101.1$ & $   0.508\pm   0.023$ & $  -0.514\pm   0.034$ & $  41.6$\\
091029   & $ -43.1$ & $  82.2$ & $   0.684\pm   0.009$ & $  -0.641\pm   0.021$ & $  39.2$\\
091102   & $  -6.9$ & $  13.0$ & $  -0.088\pm   0.025$ & $  -0.537\pm   0.046$ & $   6.6$\\
091127   & $ -10.5$ & $  20.1$ & $   1.186\pm   0.011$ & $   0.844\pm   0.030$ & $   7.1$\\
091130B  & $-117.4$ & $ 227.2$ & $   0.522\pm   0.024$ & $  -0.868\pm   0.038$ & $ 112.5$\\
091208A  & $ -30.8$ & $  61.3$ & $   0.329\pm   0.020$ & $  -0.613\pm   0.023$ & $  29.1$\\
091208B  & $ -10.4$ & $  21.6$ & $   0.678\pm   0.019$ & $   0.368\pm   0.038$ & $  14.9$\\
091221   & $-119.0$ & $ 118.2$ & $   1.002\pm   0.009$ & $  -0.407\pm   0.020$ & $  68.5$\\
100111A  & $ -15.3$ & $  16.2$ & $   0.087\pm   0.023$ & $  -0.630\pm   0.033$ & $  12.9$\\
100119A  & $ -48.4$ & $  47.6$ & $   1.279\pm   0.012$ & $  -0.030\pm   0.025$ & $  23.8$\\
100212A  & $-231.7$ & $ 316.8$ & $   0.359\pm   0.048$ & $  -0.430\pm   0.036$ & $ 136.0$\\
100413A  & $-207.1$ & $ 422.4$ & $   0.958\pm   0.016$ & $  -0.754\pm   0.091$ & $ 191.0$\\
100423A  & $-118.4$ & $ 156.0$ & $   1.120\pm   0.008$ & $   0.179\pm   0.030$ & $ 160.0$\\
100425A  & $ -39.9$ & $  74.0$ & $   0.108\pm   0.050$ & $  -0.775\pm   0.046$ & $  37.0$\\
100504A  & $-116.3$ & $ 210.3$ & $   0.620\pm   0.022$ & $  -0.704\pm   0.029$ & $  97.3$\\
100522A  & $ -35.8$ & $  71.1$ & $   0.576\pm   0.015$ & $   0.076\pm   0.032$ & $  35.3$\\
100606A  & $ -75.1$ & $ 150.3$ & $   0.888\pm   0.016$ & $  -0.324\pm   0.044$ & $ 480.0$\\
100615A  & $ -47.1$ & $  94.4$ & $   0.962\pm   0.006$ & $  -0.149\pm   0.019$ & $  39.0$\\
100619A  & $-108.5$ & $ 212.6$ & $   0.883\pm   0.011$ & $  -0.240\pm   0.015$ & $  97.5$\\
100621A  & $-103.6$ & $ 190.5$ & $   1.580\pm   0.002$ & $   0.202\pm   0.008$ & $  63.6$\\
100704A  & $-217.4$ & $ 397.4$ & $   0.992\pm   0.011$ & $  -0.251\pm   0.017$ & $ 197.5$\\
100725A  & $ -86.5$ & $ 168.8$ & $   0.512\pm   0.019$ & $  -0.999\pm   0.052$ & $ 141.0$\\
100725B  & $-222.9$ & $ 436.2$ & $   1.078\pm   0.012$ & $  -0.525\pm   0.022$ & $ 200.0$\\
100727A  & $-134.8$ & $ 102.7$ & $   0.375\pm   0.024$ & $  -0.811\pm   0.034$ & $  84.0$\\
100728A  & $-110.0$ & $ 723.0$ & $   1.736\pm   0.002$ & $  -0.112\pm   0.024$ & $ 198.5$\\
100802A  & $-200.0$ & $ 936.7$ & $   0.836\pm   0.019$ & $  -0.972\pm   0.033$ & $ 487.0$\\
100814A  & $-235.0$ & $ 460.6$ & $   1.154\pm   0.008$ & $  -0.484\pm   0.013$ & $ 174.5$\\
100823A  & $ -20.7$ & $  38.8$ & $  -0.087\pm   0.038$ & $  -0.949\pm   0.042$ & $  16.9$\\
100902A  & $ -89.9$ & $ 859.1$ & $   0.786\pm   0.024$ & $  -0.984\pm   0.029$ & $ 428.8$\\
100906A  & $-130.6$ & $ 261.0$ & $   1.283\pm   0.005$ & $   0.104\pm   0.016$ & $ 114.4$\\
100924A  & $ -79.4$ & $ 155.5$ & $   1.116\pm   0.010$ & $  -0.192\pm   0.032$ & $  96.0$\\
101008A  & $ -10.5$ & $  15.5$ & $   0.051\pm   0.028$ & $  -0.494\pm   0.063$ & $ 104.0$\\
101011A  & $ -36.9$ & $  73.1$ & $   0.307\pm   0.023$ & $  -0.701\pm   0.041$ & $  71.5$\\
101017A  & $-116.9$ & $ 217.5$ & $   1.416\pm   0.004$ & $   0.176\pm   0.035$ & $  70.0$\\
101023A  & $-119.9$ & $ 225.7$ & $   1.630\pm   0.005$ & $   0.450\pm   0.008$ & $  80.8$\\
101024A  & $ -20.1$ & $  39.7$ & $   0.423\pm   0.014$ & $   0.005\pm   0.024$ & $  18.7$\\
101030A  & $-183.9$ & $ 148.7$ & $   0.733\pm   0.021$ & $  -0.948\pm   0.027$ & $  92.0$\\
101117B  & $  -3.8$ & $   7.6$ & $   0.232\pm   0.015$ & $  -0.082\pm   0.037$ & $   5.2$\\
110102A  & $-100.0$ & $ 617.4$ & $   1.417\pm   0.005$ & $   0.074\pm   0.017$ & $ 264.0$\\
110106B  & $ -26.0$ & $  44.7$ & $   0.544\pm   0.019$ & $  -0.484\pm   0.049$ & $  24.8$\\
110119A  & $-230.7$ & $ 438.5$ & $   1.059\pm   0.009$ & $  -0.397\pm   0.030$ & $ 208.0$\\
110201A  & $ -10.1$ & $  19.4$ & $  -0.019\pm   0.021$ & $  -0.742\pm   0.043$ & $  13.0$\\
110205A  & $-239.0$ & $ 727.0$ & $   1.439\pm   0.005$ & $  -0.279\pm   0.030$ & $ 257.0$\\
110207A  & $ -70.6$ & $ 141.2$ & $   0.392\pm   0.025$ & $  -0.219\pm   0.041$ & $  80.3$\\
110213A  & $ -53.8$ & $  48.0$ & $   0.979\pm   0.018$ & $   0.092\pm   0.033$ & $  48.0$\\
110315A  & $-161.8$ & $ 127.0$ & $   0.846\pm   0.015$ & $  -0.685\pm   0.048$ & $  77.0$\\
110318A  & $ -24.3$ & $  27.1$ & $   0.892\pm   0.010$ & $   0.002\pm   0.017$ & $  16.0$\\
110319A  & $ -23.9$ & $  47.8$ & $   0.548\pm   0.010$ & $  -0.592\pm   0.019$ & $  19.3$\\
110402A  & $ -31.6$ & $  74.2$ & $   0.709\pm   0.030$ & $   0.312\pm   0.060$ & $  60.9$\\
110411A  & $-106.8$ & $ 182.8$ & $   0.826\pm   0.012$ & $  -0.819\pm   0.026$ & $  80.3$\\
110414A  & $-205.0$ & $ 304.8$ & $   0.758\pm   0.031$ & $  -0.997\pm   0.039$ & $ 152.0$\\
110420A  & $ -24.6$ & $  50.1$ & $   1.043\pm   0.009$ & $   0.174\pm   0.018$ & $  11.8$\\
110422A  & $ -55.3$ & $  77.2$ & $   1.738\pm   0.003$ & $   0.590\pm   0.013$ & $  25.9$\\
110503A  & $ -11.1$ & $  17.0$ & $   1.143\pm   0.009$ & $   0.579\pm   0.021$ & $  10.0$\\
110519A  & $ -34.9$ & $  58.2$ & $   0.877\pm   0.009$ & $  -0.274\pm   0.016$ & $  27.2$\\
110610A  & $ -76.6$ & $ 123.0$ & $   0.882\pm   0.009$ & $  -0.270\pm   0.022$ & $  46.4$\\
110625A  & $-157.3$ & $ 280.8$ & $   1.641\pm   0.008$ & $   0.828\pm   0.025$ & $  44.5$\\
110709A  & $ -64.0$ & $ 116.9$ & $   1.187\pm   0.004$ & $  -0.046\pm   0.027$ & $  44.7$\\
110715A  & $ -19.2$ & $  35.0$ & $   1.266\pm   0.003$ & $   0.881\pm   0.009$ & $  13.0$\\
110731A  & $ -26.4$ & $  49.7$ & $   0.915\pm   0.006$ & $   0.245\pm   0.016$ & $  38.8$\\
110801A  & $-239.2$ & $ 792.5$ & $   0.976\pm   0.017$ & $  -0.813\pm   0.056$ & $ 385.0$\\
\enddata
%% Text for table notes should follow after the \enddata but before
%% the \end{deluxetable}. Make sure there is at least one \tablenotemark
%% in the table for each \tablenotetext.
\tablecomments{The PDS is calculated in the time interval reported.}
\tablenotetext{a}{Referred to the BAT trigger time, and calculated in the observer frame.}
%\tablenotetext{b}{Selection from the $z$-silver sample.}
\end{deluxetable}
\twocolumn

%%%%%%%%%%% z silver sample (97 GRBs) %%%%%%%%%%%%%%%%%%%
\onecolumn
\begin{deluxetable}{lrrrrr}
\tablecolumns{6}
\tabletypesize{\scriptsize}
%\rotate
\tablecaption{Table of the $z$-silver sample including 97 GRBs.
\label{tbl-2}}
\tablewidth{0pt}
\tablehead{
\colhead{GRB} & \colhead{$z$} & \colhead{$t_{\rm start}$\tablenotemark{a}} & \colhead{$t_{\rm stop}$\tablenotemark{a}} & \colhead{Log(Count Fluence)} & \colhead{Log(peak rate)\tablenotemark{b}}\\
 & & (s) & (s) & \colhead{(count~det$^{-1}$)} & \colhead{(count~s$^{-1}$~det$^{-1}$)}}
\startdata
050315   & $ 1.949$ & $ -61.0$ & $  65.0$ & $   0.785\pm   0.014$ & $  -0.226\pm   0.032$\\
050318   & $ 1.440$ & $ -13.4$ & $  13.3$ & $   0.375\pm   0.015$ & $   0.051\pm   0.023$\\
050319   & $ 3.240$ & $ -37.7$ & $  44.1$ & $   0.396\pm   0.029$ & $  -0.186\pm   0.031$\\
050401   & $ 2.900$ & $ -11.1$ & $  16.6$ & $   1.094\pm   0.014$ & $   0.739\pm   0.031$\\
050505   & $ 4.270$ & $ -12.1$ & $  18.5$ & $   0.549\pm   0.027$ & $  -0.008\pm   0.037$\\
050525A  & $ 0.606$ & $  -7.7$ & $  15.6$ & $   1.413\pm   0.001$ & $   1.042\pm   0.007$\\
050603   & $ 2.821$ & $  -6.9$ & $  10.9$ & $   1.062\pm   0.013$ & $   1.360\pm   0.021$\\
050730   & $ 3.967$ & $ -42.4$ & $  57.6$ & $   0.673\pm   0.019$ & $  -0.565\pm   0.032$\\
050820A  & $ 2.612$ & $ -68.0$ & $  83.0$ & $   0.763\pm   0.017$ & $   0.021\pm   0.021$\\
050922C  & $ 2.198$ & $  -2.5$ & $   2.9$ & $   0.405\pm   0.009$ & $   0.507\pm   0.020$\\
051111   & $ 1.550$ & $ -23.8$ & $  40.4$ & $   0.789\pm   0.010$ & $  -0.087\pm   0.018$\\
060115   & $ 3.530$ & $ -26.1$ & $  49.5$ & $   0.543\pm   0.020$ & $  -0.372\pm   0.028$\\
060206   & $ 4.048$ & $  -2.3$ & $   4.3$ & $   0.201\pm   0.013$ & $   0.273\pm   0.016$\\
060210   & $ 3.910$ & $ -61.1$ & $  61.6$ & $   1.082\pm   0.014$ & $   0.212\pm   0.031$\\
060223A  & $ 4.410$ & $  -2.6$ & $   4.1$ & $   0.057\pm   0.025$ & $  -0.069\pm   0.033$\\
060418   & $ 1.489$ & $ -68.9$ & $ 114.8$ & $   1.149\pm   0.009$ & $   0.414\pm   0.025$\\
060502A  & $ 1.510$ & $ -11.6$ & $  16.5$ & $   0.547\pm   0.013$ & $  -0.300\pm   0.020$\\
060510B  & $ 4.900$ & $ -40.5$ & $  71.5$ & $   0.886\pm   0.012$ & $  -0.549\pm   0.027$\\
060526   & $ 3.210$ & $ -56.8$ & $ 123.0$ & $   0.415\pm   0.041$ & $  -0.070\pm   0.036$\\
060607A  & $ 3.082$ & $ -30.9$ & $  52.5$ & $   0.658\pm   0.014$ & $  -0.154\pm   0.018$\\
060614   & $ 0.125$ & $-147.7$ & $ 291.3$ & $   1.596\pm   0.004$ & $   0.457\pm   0.035$\\
060707   & $ 3.430$ & $ -25.9$ & $  19.5$ & $   0.482\pm   0.026$ & $  -0.295\pm   0.042$\\
060714   & $ 2.710$ & $ -36.2$ & $  62.9$ & $   0.728\pm   0.020$ & $  -0.207\pm   0.036$\\
060814   & $ 0.840$ & $ -40.7$ & $ 209.2$ & $   1.379\pm   0.004$ & $   0.225\pm   0.014$\\
060904B  & $ 0.703$ & $-102.0$ & $ 203.7$ & $   0.429\pm   0.036$ & $  -0.262\pm   0.027$\\
060906   & $ 3.685$ & $ -10.2$ & $  11.7$ & $   0.644\pm   0.017$ & $  -0.062\pm   0.040$\\
060908   & $ 1.884$ & $ -10.2$ & $  11.2$ & $   0.601\pm   0.011$ & $   0.110\pm   0.029$\\
060912A  & $ 0.937$ & $  -2.8$ & $   5.3$ & $   0.373\pm   0.012$ & $   0.332\pm   0.021$\\
060927   & $ 5.467$ & $  -3.8$ & $   7.3$ & $   0.319\pm   0.016$ & $   0.352\pm   0.024$\\
061007   & $ 1.262$ & $ -17.7$ & $ 158.2$ & $   1.789\pm   0.002$ & $   0.696\pm   0.013$\\
061021   & $ 0.346$ & $ -40.6$ & $  81.1$ & $   0.671\pm   0.011$ & $   0.061\pm   0.018$\\
061121   & $ 1.314$ & $ -58.4$ & $ 115.9$ & $   1.366\pm   0.003$ & $   0.865\pm   0.009$\\
061126   & $ 1.159$ & $ -39.9$ & $  72.6$ & $   0.988\pm   0.009$ & $   0.456\pm   0.016$\\
061222A  & $ 2.088$ & $ -36.8$ & $  71.1$ & $   1.124\pm   0.006$ & $   0.661\pm   0.012$\\
070306   & $ 1.496$ & $ -95.8$ & $ 195.4$ & $   0.993\pm   0.014$ & $   0.092\pm   0.017$\\
070318   & $ 0.836$ & $ -34.8$ & $  68.4$ & $   0.623\pm   0.012$ & $  -0.388\pm   0.021$\\
070411   & $ 2.954$ & $ -31.4$ & $  50.0$ & $   0.698\pm   0.015$ & $  -0.379\pm   0.029$\\
070529   & $ 2.500$ & $ -33.1$ & $  67.9$ & $   0.605\pm   0.034$ & $  -0.195\pm   0.071$\\
070612A  & $ 0.617$ & $-147.8$ & $ 318.0$ & $   1.225\pm   0.017$ & $  -0.611\pm   0.031$\\
070721B  & $ 3.626$ & $ -51.8$ & $ 117.3$ & $   0.745\pm   0.025$ & $  -0.081\pm   0.034$\\
071003   & $ 1.604$ & $ -70.1$ & $ 131.6$ & $   1.030\pm   0.013$ & $   0.255\pm   0.021$\\
071010B  & $ 0.947$ & $ -38.8$ & $  39.6$ & $   0.862\pm   0.007$ & $   0.166\pm   0.011$\\
071020   & $ 2.145$ & $  -4.1$ & $   5.4$ & $   0.499\pm   0.007$ & $   0.575\pm   0.015$\\
071117   & $ 1.331$ & $  -2.2$ & $   4.0$ & $   0.409\pm   0.010$ & $   0.403\pm   0.018$\\
080210   & $ 2.641$ & $  -9.6$ & $  10.9$ & $   0.414\pm   0.017$ & $  -0.272\pm   0.030$\\
080310   & $ 2.430$ & $ -24.5$ & $ 192.3$ & $   0.677\pm   0.024$ & $  -0.441\pm   0.037$\\
080319B  & $ 0.937$ & $ -35.9$ & $  32.1$ & $   2.046\pm   0.001$ & $   0.876\pm   0.008$\\
080319C  & $ 1.950$ & $  -4.9$ & $   9.3$ & $   0.579\pm   0.012$ & $   0.244\pm   0.028$\\
080330   & $ 1.510$ & $ -23.3$ & $  46.7$ & $  -0.130\pm   0.074$ & $  -0.567\pm   0.060$\\
080411   & $ 1.030$ & $ -38.6$ & $  76.8$ & $   1.643\pm   0.002$ & $   1.047\pm   0.007$\\
080413A  & $ 2.433$ & $ -14.0$ & $  28.4$ & $   0.729\pm   0.010$ & $   0.355\pm   0.013$\\
080413B  & $ 1.100$ & $  -3.3$ & $   5.2$ & $   0.624\pm   0.011$ & $   0.586\pm   0.022$\\
080430   & $ 0.767$ & $ -11.5$ & $  22.7$ & $   0.370\pm   0.013$ & $  -0.235\pm   0.018$\\
080603B  & $ 2.690$ & $ -18.3$ & $  36.7$ & $   0.635\pm   0.012$ & $   0.257\pm   0.023$\\
080605   & $ 1.640$ & $ -14.9$ & $  23.9$ & $   1.290\pm   0.004$ & $   0.852\pm   0.015$\\
080607   & $ 3.036$ & $ -38.9$ & $  72.4$ & $   1.544\pm   0.005$ & $   1.142\pm   0.020$\\
080721   & $ 2.591$ & $  -9.6$ & $  10.9$ & $   1.246\pm   0.013$ & $   1.014\pm   0.040$\\
080804   & $ 2.204$ & $ -10.4$ & $  21.4$ & $   0.699\pm   0.018$ & $  -0.028\pm   0.027$\\
080805   & $ 1.505$ & $ -35.4$ & $  68.0$ & $   0.641\pm   0.013$ & $  -0.481\pm   0.023$\\
080810   & $ 3.350$ & $ -28.2$ & $  50.6$ & $   0.854\pm   0.012$ & $   0.052\pm   0.037$\\
080905B  & $ 2.374$ & $ -28.5$ & $  56.9$ & $   0.487\pm   0.030$ & $  -0.241\pm   0.043$\\
080906   & $ 2.130$ & $ -76.5$ & $  85.4$ & $   0.832\pm   0.014$ & $  -0.437\pm   0.022$\\
080916A  & $ 0.689$ & $ -45.3$ & $  86.6$ & $   0.837\pm   0.008$ & $  -0.240\pm   0.018$\\
080928   & $ 1.692$ & $ -48.9$ & $ 148.6$ & $   0.680\pm   0.019$ & $  -0.076\pm   0.038$\\
081008   & $ 1.968$ & $ -80.7$ & $ 114.0$ & $   0.868\pm   0.013$ & $  -0.335\pm   0.025$\\
081028   & $ 3.038$ & $ -38.9$ & $ 114.9$ & $   0.829\pm   0.014$ & $  -0.696\pm   0.026$\\
081222   & $ 2.770$ & $  -9.8$ & $  19.3$ & $   0.917\pm   0.005$ & $   0.586\pm   0.009$\\
090102   & $ 1.547$ & $  -4.9$ & $  33.4$ & $   0.988\pm   0.017$ & $   0.404\pm   0.047$\\
090418A  & $ 1.608$ & $ -26.7$ & $  47.8$ & $   0.849\pm   0.013$ & $  -0.180\pm   0.034$\\
090423   & $ 8.100$ & $  -1.2$ & $   2.5$ & $   0.098\pm   0.019$ & $   0.272\pm   0.030$\\
090424   & $ 0.544$ & $ -53.9$ & $ 106.5$ & $   1.515\pm   0.005$ & $   1.140\pm   0.012$\\
090516   & $ 4.109$ & $ -45.3$ & $  87.5$ & $   1.114\pm   0.029$ & $   0.109\pm   0.058$\\
090530   & $ 1.280$ & $ -18.0$ & $  35.8$ & $   0.245\pm   0.033$ & $   0.017\pm   0.035$\\
090618   & $ 0.540$ & $-139.4$ & $ 267.7$ & $   2.239\pm   0.001$ & $   0.841\pm   0.004$\\
090715B  & $ 3.000$ & $ -59.9$ & $ 137.5$ & $   0.984\pm   0.012$ & $   0.278\pm   0.012$\\
090812   & $ 2.452$ & $ -28.8$ & $  53.3$ & $   0.934\pm   0.009$ & $   0.261\pm   0.027$\\
090926B  & $ 1.240$ & $ -76.5$ & $ 125.6$ & $   1.033\pm   0.011$ & $  -0.128\pm   0.020$\\
091018   & $ 0.971$ & $  -3.5$ & $   6.8$ & $   0.493\pm   0.007$ & $   0.378\pm   0.013$\\
091020   & $ 1.710$ & $ -17.4$ & $  28.6$ & $   0.754\pm   0.011$ & $   0.115\pm   0.017$\\
091024   & $ 1.092$ & $ -56.0$ & $  92.2$ & $   0.962\pm   0.014$ & $  -0.282\pm   0.028$\\
091029   & $ 2.752$ & $ -11.5$ & $  21.9$ & $   0.681\pm   0.009$ & $  -0.086\pm   0.018$\\
091127   & $ 0.490$ & $  -7.0$ & $  13.5$ & $   1.178\pm   0.011$ & $   0.991\pm   0.025$\\
091208B  & $ 1.063$ & $  -5.1$ & $  10.5$ & $   0.671\pm   0.018$ & $   0.670\pm   0.037$\\
100425A  & $ 1.755$ & $ -14.5$ & $  26.9$ & $   0.092\pm   0.049$ & $  -0.330\pm   0.048$\\
100621A  & $ 0.542$ & $ -67.2$ & $ 123.5$ & $   1.578\pm   0.002$ & $   0.389\pm   0.008$\\
100814A  & $ 1.440$ & $ -96.3$ & $ 188.8$ & $   1.152\pm   0.007$ & $  -0.121\pm   0.011$\\
100906A  & $ 1.727$ & $ -47.9$ & $  95.7$ & $   1.280\pm   0.005$ & $   0.524\pm   0.016$\\
101213A  & $ 0.414$ & $ -51.2$ & $  96.8$ & $   0.860\pm   0.014$ & $  -0.464\pm   0.036$\\
110106B  & $ 0.618$ & $ -16.1$ & $  27.6$ & $   0.538\pm   0.018$ & $  -0.274\pm   0.047$\\
110205A  & $ 2.200$ & $ -74.7$ & $ 227.2$ & $   1.434\pm   0.005$ & $   0.190\pm   0.024$\\
110213A  & $ 1.460$ & $ -21.9$ & $  19.5$ & $   0.977\pm   0.017$ & $   0.479\pm   0.034$\\
110422A  & $ 1.770$ & $ -20.0$ & $  27.9$ & $   1.731\pm   0.003$ & $   0.995\pm   0.010$\\
110503A  & $ 1.613$ & $  -4.2$ & $   6.5$ & $   1.138\pm   0.008$ & $   0.985\pm   0.018$\\
110715A  & $ 0.820$ & $ -10.6$ & $  19.2$ & $   1.262\pm   0.003$ & $   1.129\pm   0.009$\\
110731A  & $ 2.830$ & $  -6.9$ & $  13.0$ & $   0.909\pm   0.005$ & $   0.793\pm   0.012$\\
110801A  & $ 1.858$ & $ -83.7$ & $ 277.3$ & $   0.963\pm   0.017$ & $  -0.382\pm   0.042$\\
110818A  & $ 3.360$ & $ -15.9$ & $  29.9$ & $   0.749\pm   0.020$ & $  -0.148\pm   0.045$\\
\enddata
%% Text for table notes should follow after the \enddata but before
%% the \end{deluxetable}. Make sure there is at least one \tablenotemark
%% in the table for each \tablenotetext.
\tablecomments{The PDS is calculated in the time interval reported.}
\tablenotetext{a}{Referred to the BAT trigger time, and calculated in the source rest frame.}
\tablenotetext{b}{Calculated in the source rest frame.}
\end{deluxetable}
\twocolumn

%%%%%%%%%%% z golden sample (49 GRBs) %%%%%%%%%%%%%%%%%%%
\onecolumn
\begin{deluxetable}{lrrrrr}
\tablecolumns{6}
\tabletypesize{\scriptsize}
%\rotate
\tablecaption{Table of the $z$-golden sample including 49 GRBs.
\label{tbl-3}}
\tablewidth{0pt}
\tablehead{
\colhead{GRB} & \colhead{$z$} & \colhead{$t_{\rm start}$\tablenotemark{a}} & \colhead{$t_{\rm stop}$\tablenotemark{a}} & \colhead{Log(Count Fluence)} & \colhead{Log(peak rate)\tablenotemark{b}}\\
 & & (s) & (s) & \colhead{(count~det$^{-1}$)} & \colhead{(count~s$^{-1}$~det$^{-1}$)}}
\startdata
050315   & $ 1.949$ & $ -61.0$ & $  65.0$ & $   0.631\pm   0.017$ & $  -0.367\pm   0.036$\\
050319   & $ 3.240$ & $ -41.1$ & $  44.1$ & $   0.353\pm   0.031$ & $  -0.213\pm   0.042$\\
050401   & $ 2.900$ & $ -11.1$ & $  16.6$ & $   1.027\pm   0.015$ & $   0.677\pm   0.033$\\
050603   & $ 2.821$ & $  -6.9$ & $  10.9$ & $   0.997\pm   0.014$ & $   1.212\pm   0.024$\\
050922C  & $ 2.198$ & $  -2.5$ & $   2.9$ & $   0.332\pm   0.010$ & $   0.471\pm   0.020$\\
051111   & $ 1.550$ & $ -23.8$ & $  40.4$ & $   0.655\pm   0.011$ & $  -0.172\pm   0.021$\\
060418   & $ 1.489$ & $ -68.9$ & $ 114.8$ & $   0.959\pm   0.010$ & $   0.263\pm   0.029$\\
060502A  & $ 1.510$ & $ -11.6$ & $  16.5$ & $   0.400\pm   0.014$ & $  -0.353\pm   0.029$\\
060526   & $ 3.210$ & $ -56.8$ & $ 123.0$ & $   0.386\pm   0.040$ & $  -0.113\pm   0.038$\\
060607A  & $ 3.082$ & $ -30.9$ & $  52.5$ & $   0.612\pm   0.014$ & $  -0.210\pm   0.017$\\
060707   & $ 3.430$ & $ -25.9$ & $  19.5$ & $   0.438\pm   0.027$ & $  -0.339\pm   0.042$\\
060714   & $ 2.710$ & $ -36.2$ & $  62.9$ & $   0.665\pm   0.021$ & $  -0.271\pm   0.039$\\
060908   & $ 1.884$ & $ -10.2$ & $  11.2$ & $   0.505\pm   0.012$ & $   0.044\pm   0.031$\\
061222A  & $ 2.088$ & $ -36.8$ & $  71.1$ & $   1.029\pm   0.006$ & $   0.592\pm   0.013$\\
070306   & $ 1.496$ & $ -95.8$ & $ 195.4$ & $   0.800\pm   0.018$ & $  -0.090\pm   0.021$\\
070411   & $ 2.954$ & $ -31.4$ & $  50.0$ & $   0.631\pm   0.016$ & $  -0.363\pm   0.036$\\
070529   & $ 2.500$ & $ -33.1$ & $  67.9$ & $   0.561\pm   0.033$ & $  -0.250\pm   0.073$\\
071003   & $ 1.604$ & $ -70.1$ & $ 131.6$ & $   0.887\pm   0.015$ & $   0.152\pm   0.022$\\
071020   & $ 2.145$ & $  -4.1$ & $   5.4$ & $   0.415\pm   0.008$ & $   0.502\pm   0.015$\\
080210   & $ 2.641$ & $  -9.6$ & $  10.9$ & $   0.340\pm   0.018$ & $  -0.323\pm   0.030$\\
080310   & $ 2.430$ & $ -20.0$ & $ 192.3$ & $   0.579\pm   0.025$ & $  -0.532\pm   0.035$\\
080319C  & $ 1.950$ & $  -4.9$ & $   9.3$ & $   0.497\pm   0.013$ & $   0.139\pm   0.033$\\
080413A  & $ 2.433$ & $ -14.0$ & $  28.4$ & $   0.646\pm   0.011$ & $   0.295\pm   0.014$\\
080603B  & $ 2.690$ & $ -18.3$ & $  36.7$ & $   0.569\pm   0.012$ & $   0.155\pm   0.018$\\
080605   & $ 1.640$ & $ -14.9$ & $  23.9$ & $   1.166\pm   0.004$ & $   0.782\pm   0.016$\\
080607   & $ 3.036$ & $ -38.9$ & $  72.4$ & $   1.472\pm   0.006$ & $   1.056\pm   0.022$\\
080721   & $ 2.591$ & $  -9.6$ & $  10.9$ & $   1.162\pm   0.014$ & $   0.879\pm   0.033$\\
080804   & $ 2.204$ & $ -10.4$ & $  21.4$ & $   0.621\pm   0.018$ & $  -0.047\pm   0.029$\\
080805   & $ 1.505$ & $ -35.4$ & $  68.0$ & $   0.440\pm   0.017$ & $  -0.530\pm   0.026$\\
080810   & $ 3.350$ & $ -28.2$ & $  50.6$ & $   0.795\pm   0.012$ & $   0.000\pm   0.038$\\
080905B  & $ 2.374$ & $ -28.5$ & $  56.9$ & $   0.395\pm   0.032$ & $  -0.310\pm   0.042$\\
080906   & $ 2.130$ & $ -76.5$ & $  85.4$ & $   0.740\pm   0.015$ & $  -0.501\pm   0.026$\\
080928   & $ 1.692$ & $ -48.9$ & $ 150.0$ & $   0.511\pm   0.023$ & $  -0.225\pm   0.031$\\
081008   & $ 1.968$ & $ -80.7$ & $ 114.0$ & $   0.747\pm   0.014$ & $  -0.408\pm   0.030$\\
081222   & $ 2.770$ & $  -9.8$ & $  19.3$ & $   0.857\pm   0.005$ & $   0.523\pm   0.014$\\
090102   & $ 1.547$ & $  -4.9$ & $  33.4$ & $   0.865\pm   0.019$ & $   0.290\pm   0.053$\\
090418A  & $ 1.608$ & $ -26.7$ & $  47.8$ & $   0.721\pm   0.014$ & $  -0.252\pm   0.046$\\
090715B  & $ 3.000$ & $ -59.9$ & $ 140.0$ & $   0.923\pm   0.012$ & $   0.221\pm   0.013$\\
090812   & $ 2.452$ & $ -28.8$ & $  53.3$ & $   0.862\pm   0.009$ & $   0.205\pm   0.027$\\
091020   & $ 1.710$ & $ -17.4$ & $  28.6$ & $   0.602\pm   0.013$ & $   0.007\pm   0.019$\\
091029   & $ 2.752$ & $ -11.5$ & $  21.9$ & $   0.612\pm   0.009$ & $  -0.143\pm   0.023$\\
100814A  & $ 1.440$ & $ -96.3$ & $ 188.8$ & $   0.976\pm   0.008$ & $  -0.194\pm   0.012$\\
100906A  & $ 1.727$ & $ -47.9$ & $  95.7$ & $   1.118\pm   0.006$ & $   0.415\pm   0.017$\\
110205A  & $ 2.200$ & $ -74.7$ & $ 227.2$ & $   1.331\pm   0.006$ & $   0.105\pm   0.026$\\
110213A  & $ 1.460$ & $ -21.9$ & $  19.5$ & $   0.770\pm   0.022$ & $   0.321\pm   0.040$\\
110422A  & $ 1.770$ & $ -20.0$ & $  27.9$ & $   1.634\pm   0.003$ & $   0.939\pm   0.013$\\
110503A  & $ 1.613$ & $  -4.2$ & $   6.5$ & $   1.028\pm   0.009$ & $   0.893\pm   0.020$\\
110731A  & $ 2.830$ & $  -6.9$ & $  13.0$ & $   0.841\pm   0.006$ & $   0.740\pm   0.013$\\
110801A  & $ 1.858$ & $ -83.7$ & $ 277.3$ & $   0.804\pm   0.020$ & $  -0.555\pm   0.044$\\
\enddata
%% Text for table notes should follow after the \enddata but before
%% the \end{deluxetable}. Make sure there is at least one \tablenotemark
%% in the table for each \tablenotetext.
\tablecomments{The PDS is calculated in the time interval reported.}
\tablenotetext{a}{Referred to the BAT trigger time, and calculated in the source rest frame.}
\tablenotetext{b}{Calculated in the source rest frame.}
\end{deluxetable}
\twocolumn

%%%%%%%%%%% Best-fit parameters %%%%%%%%%%%%%%%%%%%
\onecolumn
\begin{deluxetable}{lrccccccccc}
\tablecolumns{11}
\tabletypesize{\scriptsize}
%\rotate
\tablecaption{Best fit parameters of the average PDS for different samples of GRBs
\label{tbl-4}}
\tablewidth{0pt}
\tablehead{
\colhead{} & \colhead{} & \multicolumn{4}{c}{Variance norm.} & \colhead{} & \multicolumn{4}{c}{Peak norm.} \\
\cline{3-6} \cline{8-11} \\ 
\colhead{Sample} & \colhead{Size} &
\colhead{$\alpha_1$} & \colhead{$f_{\rm b}$} & \colhead{$\alpha_2$} & \colhead{$\chi^2$/dof} & \colhead{} &
\colhead{$\alpha_1$} & \colhead{$f_{\rm b}$} & \colhead{$\alpha_2$} & \colhead{$\chi^2$/dof}\\
 & & & \colhead{($10^{-2}$~Hz)} & & & & & \colhead{($10^{-2}$~Hz)} & &}
\startdata
full       & 244 & $1.03_{-0.05}^{+0.05}$ & $ 3.0_{-0.4}^{+0.5}$ & $1.73_{-0.03}^{+0.04}$ & $73.5/63$ 
           &     & $1.25_{-0.12}^{+0.11}$ & $ 2.0_{-0.4}^{+0.6}$ & $1.90_{-0.06}^{+0.07}$ & $15.4/21$\\
$z$-silver & 97  & $1.01_{-0.08}^{+0.07}$ & $ 6.9_{-1.2}^{+1.5}$ & $1.77_{-0.05}^{+0.05}$ & $60.4/82$
           &     & $1.38_{-0.20}^{+0.14}$ & $ 6.8_{-2.7}^{+3.2}$ & $2.06_{-0.18}^{+0.25}$ & $15.3/25$\\
$z$-golden (RF) & 49  & $0.86_{-0.15}^{+0.14}$ & $ 5.3_{-0.9}^{+1.2}$ & $1.71_{-0.05}^{+0.05}$ & $39.0/66$
                &     & $1.02_{-0.24}^{+0.19}$ & $ 4.6_{-1.1}^{+1.6}$ & $1.83_{-0.08}^{+0.09}$ & $19.7/39$\\
$z$-golden (OF) & 49  & $0.95_{-0.11}^{+0.09}$ & $ 2.4_{-0.4}^{+0.5}$ & $1.77_{-0.05}^{+0.06}$ & $40.1/40$
                &     & $1.08_{-0.24}^{+0.20}$ & $ 1.8_{-0.4}^{+0.8}$ & $1.92_{-0.11}^{+0.12}$ & $14.7/17$\\
15--50~keV\tablenotemark{a} & 244 & $1.07_{-0.05}^{+0.05}$ & $ 2.8_{-0.4}^{+0.6}$ & $1.75_{-0.04}^{+0.05}$ & $58.8/46$
                            &     & $1.30_{-0.14}^{+0.13}$ & $ 2.0_{-0.5}^{+0.8}$ & $1.97_{-0.09}^{+0.10}$ & $11.2/15$\\
50--150~keV\tablenotemark{a}& 244 & $0.91_{-0.14}^{+0.11}$ & $ 2.0_{-0.5}^{+0.7}$ & $1.49_{-0.07}^{+0.08}$ & $14.0/24$
                            &     & $1.04_{-0.18}^{+0.14}$ & $ 1.8_{-0.5}^{+0.7}$ & $1.80_{-0.12}^{+0.15}$ & $6.3/14$\\
$T_{90}>80$~s\tablenotemark{a} & 90  & $1.06_{-0.07}^{+0.12}$ & $ 2.0_{-0.3}^{+0.4}$ & $1.78_{-0.06}^{+0.05}$ & $31.3/32$
                      &     & $1.18_{-0.18}^{+0.16}$ & $ 1.8_{-0.4}^{+0.8}$ & $2.00_{-0.12}^{+0.13}$ & $8.6/13$\\
$T_{90}<40$~s\tablenotemark{a}& 90  & $0.59_{-0.11}^{+0.10}$ & $ 3.2_{-0.4}^{+0.5}$ & $1.72_{-0.04}^{+0.05}$ & $93.6/59$
                      &     & $0.96_{-0.25}^{+0.20}$ & $ 3.1_{-0.5}^{+0.9}$ & $1.86_{-0.06}^{+0.06}$ & $22.1/37$\\
$P<0.4$\tablenotemark{a} & 124 & $1.03_{-0.07}^{+0.07}$ & $ 2.1_{-0.3}^{+0.4}$ & $1.73_{-0.04}^{+0.06}$ & $27.9/33$
                       &     & $1.25_{-0.15}^{+0.14}$ & $ 1.8_{-0.4}^{+0.5}$ & $2.03_{-0.10}^{+0.11}$ & $10.8/14$\\
$P>1.0$\tablenotemark{a} & 65  & $1.00_{-0.06}^{+0.06}$ & $ 6.1_{-1.0}^{+1.5}$ & $1.79_{-0.04}^{+0.04}$ & $109/115$
                        &     & $1.44_{-0.09}^{+0.06}$ & $10.9_{-3.4}^{+4.6}$ & $1.93_{-0.06}^{+0.05}$ & $89.5/117$\\
$F<4.4$\tablenotemark{a} & 97  & $0.80_{-0.11}^{+0.10}$ & $ 3.0_{-0.6}^{+1.0}$ & $1.65_{-0.05}^{+0.05}$ & $60.6/50$
                       &     & $0.94_{-0.20}^{+0.16}$ & $ 2.0_{-0.4}^{+0.6}$ & $1.79_{-0.07}^{+0.08}$ & $19.0/24$\\
$F>19.8$\tablenotemark{a} & 30  & $1.31_{-0.04}^{+0.05}$ & $ 9.4_{-1.1}^{+4.9}$ & $1.94_{-0.04}^{+0.08}$ & $120/106$
                        &     & $1.45_{-0.14}^{+0.11}$ & $ 3.6_{-0.9}^{+3.6}$ & $1.77_{-0.07}^{+0.07}$ & $19.4/44$\\
$E_{\rm iso,52}<9$\tablenotemark{b,c} & 25 & $1.04_{-0.13}^{+0.10}$ & $ 6.2_{-2.7}^{+2.5}$ & $1.70_{-0.08}^{+0.09}$ & $74.5/48$
                                    &    & $1.21_{-0.07}^{+0.07}$ & $ 8.6_{-2.3}^{+2.2}$ & $2.14_{-0.19}^{+0.23}$ & $57.6/37$\\
$E_{\rm iso,52}>21$\tablenotemark{b,c} & 25 & $1.08_{-0.17}^{+0.15}$ & $ 8.3_{-2.0}^{+3.0}$ & $1.89_{-0.07}^{+0.07}$ & $73.5/81$
                                     &    & --                   &   --               & $1.86_{-0.07}^{+0.07}$\tablenotemark{d} & $31.1/43$\tablenotemark{d}\\
 $0.1<z<1.5$\tablenotemark{b} & 32  & $1.12_{-0.08}^{+0.08}$ & $ 10.2_{-2.2}^{+2.6}$ & $1.95_{-0.11}^{+0.12}$ & $50.6/48$
                &   & --                   & --                  & $1.68_{-0.11}^{+0.10}$\tablenotemark{d} & $18.2/14$\tablenotemark{d}\\
 $2.6<z<8.1$\tablenotemark{b} & 32  & $0.67_{-0.22}^{+0.17}$ & $  6.0_{-1.3}^{+2.1}$ & $1.71_{-0.07}^{+0.08}$ & $45.0/54$
                &   & --                   & --                  & $1.69_{-0.10}^{+0.09}$\tablenotemark{d} & $17.6/28$\tablenotemark{d}\\
\enddata
%% Text for table notes should follow after the \enddata but before
%% the \end{deluxetable}. Make sure there is at least one \tablenotemark
%% in the table for each \tablenotetext.
\tablecomments{The average PDS of the samples denoted with 'RF' ('OF') refer to the source rest-frame (observer frame).
The peak count rate $p$ (count fluence $F$) is expressed in count~s$^{-1}$~det$^{-1}$ (count~det$^{-1}$)
per fully illuminated detector for an equivalent on-axis source.}
\tablenotetext{a}{Selection from the full sample.}
\tablenotetext{b}{Selection from the $z$-silver sample.}
\tablenotetext{c}{$E_{\rm iso,52} = E_{\rm iso}/10^{52}$~ergs.}
\tablenotetext{d}{Best-fit parameters refer to a simple power-law model.}
\end{deluxetable}
\twocolumn

%%%%%%%%%%% Eiso-Ep-Liso (64 GRBs) %%%%%%%%%%%%%%%%%%%
\onecolumn
\begin{deluxetable}{lrrrr}
\tablecolumns{5}
\tabletypesize{\scriptsize}
%\rotate
\tablecaption{Table of the subsample of 64 GRBs with measured $E_{\rm p,i}$, $E_{\rm iso}$, and $L_{\rm p,iso}$.
\label{tbl-5}}
\tablewidth{0pt}
\tablehead{
\colhead{GRB} & \colhead{$z$} & \colhead{Log($E_{\rm iso,52}$)\tablenotemark{a}} & \colhead{Log($E_{\rm p,i}$)\tablenotemark{b}} & \colhead{Log($L_{\rm p,iso,51}$)\tablenotemark{c}}}
\startdata
050318   & $ 1.440$ & $   0.361\pm   0.031$ & $   2.050\pm   0.096$ & $   1.036\pm   0.041$\\
050401   & $ 2.900$ & $   1.566\pm   0.086$ & $   2.657\pm   0.104$ & $   2.211\pm   0.093$\\
050525A  & $ 0.606$ & $   0.402\pm   0.076$ & $   2.117\pm   0.013$ & $   1.031\pm   0.076$\\
050603   & $ 2.821$ & $   1.806\pm   0.031$ & $   3.123\pm   0.035$ & $   3.103\pm   0.040$\\
050820A  & $ 2.612$ & $   2.013\pm   0.035$ & $   3.113\pm   0.092$ & $   2.271\pm   0.044$\\
060115   & $ 3.530$ & $   0.822\pm   0.066$ & $   2.452\pm   0.052$ & $   0.907\pm   0.074$\\
060206   & $ 4.048$ & $   0.653\pm   0.097$ & $   2.593\pm   0.051$ & $   1.725\pm   0.099$\\
060418   & $ 1.489$ & $   1.123\pm   0.088$ & $   2.743\pm   0.111$ & $   1.387\pm   0.092$\\
060526   & $ 3.210$ & $   0.436\pm   0.058$ & $   2.012\pm   0.088$ & $   0.951\pm   0.080$\\
060607A  & $ 3.080$ & $   1.065\pm   0.102$ & $   2.666\pm   0.109$ & $   1.253\pm   0.104$\\
060614   & $ 0.125$ & $  -0.702\pm   0.184$ & $   1.500\pm   0.500$ & $  -0.841\pm   0.187$\\
060707   & $ 3.425$ & $   0.755\pm   0.084$ & $   2.443\pm   0.044$ & $   0.978\pm   0.097$\\
060814   & $ 0.840$ & $   0.856\pm   0.043$ & $   2.650\pm   0.148$ & $   0.702\pm   0.045$\\
060908   & $ 1.884$ & $   1.013\pm   0.042$ & $   2.702\pm   0.087$ & $   1.522\pm   0.052$\\
060927   & $ 5.600$ & $   1.166\pm   0.063$ & $   2.675\pm   0.043$ & $   2.198\pm   0.069$\\
061007   & $ 1.261$ & $   1.952\pm   0.044$ & $   2.945\pm   0.061$ & $   1.859\pm   0.046$\\
061121   & $ 1.314$ & $   1.368\pm   0.050$ & $   3.107\pm   0.052$ & $   1.867\pm   0.051$\\
061126   & $ 1.159$ & $   1.494\pm   0.050$ & $   3.105\pm   0.138$ & $   1.961\pm   0.053$\\
061222A  & $ 2.088$ & $   1.468\pm   0.094$ & $   2.935\pm   0.075$ & $   2.004\pm   0.095$\\
071003   & $ 1.604$ & $   1.580\pm   0.052$ & $   3.313\pm   0.060$ & $   1.805\pm   0.058$\\
071010B  & $ 0.947$ & $   0.189\pm   0.239$ & $   1.996\pm   0.087$ & $   0.493\pm   0.239$\\
071020   & $ 2.145$ & $   0.951\pm   0.213$ & $   3.000\pm   0.069$ & $   2.026\pm   0.214$\\
071117   & $ 1.331$ & $   0.614\pm   0.104$ & $   2.783\pm   0.158$ & $   1.608\pm   0.106$\\
080319B  & $ 0.937$ & $   2.070\pm   0.033$ & $   3.100\pm   0.022$ & $   1.901\pm   0.034$\\
080319C  & $ 1.950$ & $   1.164\pm   0.089$ & $   2.937\pm   0.135$ & $   1.830\pm   0.094$\\
080411   & $ 1.030$ & $   1.209\pm   0.026$ & $   2.715\pm   0.058$ & $   1.613\pm   0.027$\\
080413A  & $ 2.433$ & $   0.920\pm   0.108$ & $   2.745\pm   0.138$ & $   1.546\pm   0.109$\\
080413B  & $ 1.100$ & $   0.397\pm   0.047$ & $   2.167\pm   0.088$ & $   1.359\pm   0.053$\\
080603B  & $ 2.690$ & $   1.062\pm   0.012$ & $   2.559\pm   0.118$ & $   1.684\pm   0.029$\\
080605   & $ 1.640$ & $   1.403\pm   0.031$ & $   2.811\pm   0.037$ & $   1.965\pm   0.035$\\
080607   & $ 3.036$ & $   2.300\pm   0.024$ & $   3.224\pm   0.058$ & $   2.898\pm   0.032$\\
080721   & $ 2.591$ & $   2.119\pm   0.075$ & $   3.237\pm   0.057$ & $   2.886\pm   0.086$\\
080810   & $ 3.350$ & $   1.676\pm   0.050$ & $   3.164\pm   0.053$ & $   1.874\pm   0.063$\\
080916A  & $ 0.689$ & $   0.025\pm   0.036$ & $   2.263\pm   0.043$ & $  -0.053\pm   0.041$\\
080928   & $ 1.692$ & $   0.589\pm   0.100$ & $   1.965\pm   0.107$ & $   0.833\pm   0.108$\\
081008   & $ 1.968$ & $   0.998\pm   0.040$ & $   2.408\pm   0.088$ & $   0.795\pm   0.049$\\
081028   & $ 3.038$ & $   1.261\pm   0.044$ & $   2.332\pm   0.183$ & $   0.736\pm   0.053$\\
081222   & $ 2.770$ & $   1.503\pm   0.050$ & $   2.702\pm   0.029$ & $   2.172\pm   0.051$\\
090102   & $ 1.547$ & $   1.351\pm   0.052$ & $   3.056\pm   0.063$ & $   1.767\pm   0.073$\\
090418A  & $ 1.608$ & $   1.230\pm   0.069$ & $   3.182\pm   0.109$ & $   1.200\pm   0.078$\\
090423   & $ 8.100$ & $   1.060\pm   0.138$ & $   2.652\pm   0.188$ & $   2.234\pm   0.142$\\
090424   & $ 0.544$ & $   0.660\pm   0.091$ & $   2.436\pm   0.008$ & $   1.284\pm   0.092$\\
090516   & $ 4.109$ & $   1.848\pm   0.085$ & $   2.949\pm   0.185$ & $   1.843\pm   0.107$\\
090618   & $ 0.540$ & $   1.340\pm   0.052$ & $   2.404\pm   0.070$ & $   0.943\pm   0.052$\\
090715B  & $ 3.000$ & $   1.372\pm   0.067$ & $   2.706\pm   0.144$ & $   1.666\pm   0.069$\\
090812   & $ 2.452$ & $   1.670\pm   0.076$ & $   3.273\pm   0.159$ & $   1.997\pm   0.081$\\
090926B  & $ 1.240$ & $   0.555\pm   0.022$ & $   2.309\pm   0.021$ & $   0.394\pm   0.032$\\
091018   & $ 0.971$ & $  -0.097\pm   0.044$ & $   1.710\pm   0.165$ & $   0.788\pm   0.046$\\
091020   & $ 1.710$ & $   0.957\pm   0.134$ & $   2.313\pm   0.359$ & $   0.298\pm   0.136$\\
091024   & $ 1.092$ & $   1.743\pm   0.042$ & $   2.724\pm   0.199$ & $   1.499\pm   0.052$\\
091029   & $ 2.752$ & $   0.947\pm   0.040$ & $   2.343\pm   0.128$ & $   1.180\pm   0.045$\\
091127   & $ 0.490$ & $   0.172\pm   0.051$ & $   1.731\pm   0.040$ & $   0.986\pm   0.058$\\
091208B  & $ 1.063$ & $   0.398\pm   0.031$ & $   2.404\pm   0.043$ & $   1.396\pm   0.052$\\
100621A  & $ 0.542$ & $   0.447\pm   0.055$ & $   2.159\pm   0.069$ & $   0.258\pm   0.056$\\
100814A  & $ 1.440$ & $   1.183\pm   0.051$ & $   2.410\pm   0.057$ & $   0.909\pm   0.053$\\
100906A  & $ 1.727$ & $   1.474\pm   0.040$ & $   2.455\pm   0.070$ & $   1.718\pm   0.043$\\
101213A  & $ 0.414$ & $   0.426\pm   0.085$ & $   2.604\pm   0.189$ & $   0.102\pm   0.093$\\
110205A  & $ 2.220$ & $   1.680\pm   0.058$ & $   2.841\pm   0.185$ & $   1.436\pm   0.063$\\
110213A  & $ 1.460$ & $   0.757\pm   0.061$ & $   2.325\pm   0.149$ & $   1.259\pm   0.072$\\
110422A  & $ 1.770$ & $   1.900\pm   0.045$ & $   2.622\pm   0.043$ & $   2.164\pm   0.046$\\
110503A  & $ 1.613$ & $   1.317\pm   0.038$ & $   2.739\pm   0.047$ & $   2.163\pm   0.043$\\
110715A  & $ 0.820$ & $   0.637\pm   0.044$ & $   2.341\pm   0.040$ & $   1.504\pm   0.045$\\
110731A  & $ 2.830$ & $   1.692\pm   0.040$ & $   3.065\pm   0.022$ & $   2.576\pm   0.042$\\
110818A  & $ 3.360$ & $   1.423\pm   0.045$ & $   3.037\pm   0.095$ & $   1.526\pm   0.067$\\
\enddata
%% Text for table notes should follow after the \enddata but before
%% the \end{deluxetable}. Make sure there is at least one \tablenotemark
%% in the table for each \tablenotetext.
%\tablecomments{The PDS is calculated in the time interval reported.}
\tablenotetext{a}{$E_{\rm iso,52}=E_{\rm iso}/10^{52}$~ergs.}
\tablenotetext{b}{$E_{\rm p,i}$ is expressed in keV and is measured in the source rest frame.}
\tablenotetext{c}{$L_{\rm p,iso,51}=L_{\rm p,iso}/10^{51}$~erg~s$^{-1}$.}
\end{deluxetable}
\twocolumn

\end{document}